\documentclass[aip,jcp,twocolumn]{revtex4}

\pdfoutput=1

\usepackage{graphicx}
\usepackage{bm}
\usepackage{amsmath,amssymb}
\usepackage{color}

\newcommand{\rev}[1]{\textcolor{black}{#1}}

\begin{document}

\title{Dynamical Density Functional Theory For Microswimmers}

\author{Andreas M. Menzel}
\email{menzel@thphy.uni-duesseldorf.de}
\affiliation{Institut f\"ur Theoretische Physik II, Weiche Materie,
Heinrich-Heine-Universit\"at D\"usseldorf,
40225 D\"usseldorf, Germany.}

\author{Arnab Saha}
\email{arnab@thphy.uni-duesseldorf.de}
\affiliation{Institut f\"ur Theoretische Physik II, Weiche Materie,
Heinrich-Heine-Universit\"at D\"usseldorf,
40225 D\"usseldorf, Germany.}

\author{Christian Hoell}
\affiliation{Institut f\"ur Theoretische Physik II, Weiche Materie,
Heinrich-Heine-Universit\"at D\"usseldorf,
40225 D\"usseldorf, Germany.}

\author{Hartmut L\"owen}
\email{hlowen@thphy.uni-duesseldorf.de}
\affiliation{Institut f\"ur Theoretische Physik II, Weiche Materie,
Heinrich-Heine-Universit\"at D\"usseldorf,
40225 D\"usseldorf, Germany.}

\begin{abstract}
Dynamical density functional theory (DDFT) has been successfully derived and applied to describe on the one hand passive colloidal suspensions, including hydrodynamic interactions between individual particles. On the other hand, active ``dry'' crowds of self-propelled particles have been characterized using DDFT. Here we go one essential step further and combine these two approaches. We establish a DDFT for active microswimmer suspensions. For this purpose, simple minimal model microswimmers are introduced. These microswimmers self-propel by setting the surrounding fluid into motion. They hydrodynamically interact with each other through their actively self-induced fluid flows and via the common ``passive'' hydrodynamic interactions. An effective soft steric repulsion is also taken into account. We derive the DDFT starting from common statistical approaches. Our DDFT is then tested and applied by characterizing a suspension of microswimmers the motion of which is restricted to a plane within a three-dimensional bulk fluid. Moreover, the swimmers are confined by a radially symmetric trapping potential. In certain parameter ranges, we find rotational symmetry breaking in combination with the formation of a ``hydrodynamic pumping state'', which has previously been observed in the literature as a result of particle-based simulations. An additional instability of this pumping state is revealed. 
\end{abstract}

\maketitle

\newcommand{\nwc}{\newcommand}
\nwc{\vs}{\vspace}
\nwc{\hs}{\hspace}
\nwc{\la}{\langle}
\nwc{\ra}{\rangle}
\nwc{\lw}{\linewidth}
\nwc{\nn}{\nonumber}

\nwc{\pd}[2]{\frac{\partial #1}{\partial #2}}

\section{Introduction}

Microswimmers \cite{lauga2009hydrodynamics,spagnolie2012hydrodynamics, menzel2015tuned,elgeti2015physics} are abundant in nature in the form of self-propelling microorganisms; moreover, they can be generated artificially in the laboratory. 
Prominent examples are sperm cells, usually propelling along helical paths \cite {eisenbach2006sperm}, bacteria like E.~coli moving forward by a rotational motion of their spiral-shaped flagella \cite{berg1972chemotaxis}, or synthetic Janus colloids catalyzing a chemical reaction on one of their hemispheres \cite{walther2013janus}. 

In recent years, there have been intense research activities on the individual as well as on the collective properties of such active particles \cite{lauga2009hydrodynamics,spagnolie2012hydrodynamics,romanczuk2012active, cates2012diffusive, marchetti2013hydrodynamics,menzel2015tuned,elgeti2015physics}. 
As a central difference between active systems and conventionally driven passive ones, the active systems are driven locally on the individual particle level, whereas in passive cases an external field acts on the system from outside. 
This feature, together with the interactions between active particles, 
can result in highly correlated collective motion 
and intriguing spatiotemporal patterns, see e.g.\ the transition from disordered motion to a state of collective migration \cite{vicsek1995novel, toner1995long, toner1998flocks, toner2005hydrodynamics, bertin2006boltzmann, bertin2009hydrodynamic, leoni2010swimmers}, the emergence of propagating density waves \cite{gregoire2004onset, mishra2010fluctuations, schaller2010polar, menzel2012collective, ihle2013invasion, caussin2014emergent, ohta2014soliton}, or the onset of turbulent-like behavior \cite{wensink2012meso,slomka2015generalized} and vortex formation \cite{ordemann2003pattern}. Further collective phenomena comprise dynamic clustering and motility-induced phase separation \cite{peruani2006nonequilibrium,ishikawa2008coherent, tailleur2008statistical, theurkauff2012dynamic, fily2012athermal, cates2013active, redner2013structure, palacci2013living, buttinoni2013dynamical,speck2014effective,speck2015dynamical}, crystallization \cite{bialke2012crystallization, menzel2013traveling, menzel2014active}, as well as lane formation \cite{wensink2012emergent, menzel2012soft, mccandlish2012spontaneous, menzel2013unidirectional,kogler2015lane}. Novel experimental techniques, such as automated digital tracking \cite{ballerini2008interaction, sokolov2007concentration} or the realization of active granular and artificial colloidal systems \cite{aranson2006patterns, deseigne2010collective, kudrolli2008swarming, paxton2004catalytic, tierno2010autonomously} are taking a major role in this research area. 
Often in modeling approaches, self-propulsion is implemented for ``dry'' objects by effective active forces acting on the particles \cite{hagen2014can}. In the present work, we explicitly take into account self-induced fluid flows of individual microswimmers, which they employ for propulsion. These self-induced fluid flows represent a significant contribution to the particle interactions.

Describing the collective behavior of many interacting self-propelled particles calls for statistical approaches \cite{baskaran2008enhanced,baskaran2009statistical, leoni2010swimmers,menzel2012collective,chou2012kinetic,
grossmann2015pattern,heidenreich2015hydrodynamic,chou2015active}. These comprise Boltzmann theories \cite{bertin2006boltzmann, bertin2009hydrodynamic,ihle2013invasion} and master equations  \cite{aranson2007model}. As a major benefit, it is typically relatively systematic to coarse-grain the resulting statistical equations. In this way, hydrodynamic-like equations to characterize the systems on a macroscopic level are obtained with specified expressions for the macroscopic system parameters. 
Alternatively, macroscopic equations can directly be derived from symmetry principles \cite{toner1995long, toner1998flocks, toner2005hydrodynamics,pleiner2013active,brand2014reversible}, yet leaving the expressions for the macroscopic parameters undetermined.

The statistical approach that we introduce in the following to describe suspensions of interacting active microswimmers is dynamical density functional theory (DDFT) \cite{marconi1999dynamic, marconi2000dynamic, archer2004dynamical}. It has turned out as highly effective to characterize passive systems that are determined by overdamped relaxation-type dynamics. 
Examples are spinodal decomposition \cite{archer2004dynamical}, phase separation of binary colloidal fluid mixtures \cite{archer2005dynamical}, nucleation and crystal growth \cite{van2008colloidal}, colloidal dynamics within polymeric solutions \cite{penna2003dynamic}, mixtures exposed to a temperature gradient \cite{wittkowski2012extended}, dewetting phenomena \cite{archer2010dynamical}, liquid-crystalline systems \cite{wittkowski2010derivation}, and rheology under confinement \cite{aerov2014driven,aerov2015theory}.

In the past, on the one hand, DDFT has been successfully extended for passive colloidal suspensions to include hydrodynamic interactions \cite{rex2008dynamical, rex2009dynamical,donev2014dynamic}. On the other hand, DDFT has been amended to model active self-propelled particle systems, yet by directly assigning an effective drive to the individual constituents \cite{wensink2008aggregation,wittkowski2011dynamical,menzel2013traveling,menzel2014active}. What is missing at the moment is a DDFT that \rev{brings together these two approaches and} addresses suspensions of active microswimmers. This means, a DDFT that contains active propulsion via self-induced fluid flows, including the resulting hydrodynamic interactions between the swimmers. We close this gap in the present work.

For this purpose, as a first step, a simple minimum model microswimmer must be introduced that propels via self-induced fluid flows. This step is performed in Sec.~\ref{Model}. 
Moreover, the resulting hydrodynamic and additional soft steric interactions between these swimmers are clarified, together with a confining trapping potential. 
In Sec.~\ref{Derivation}, we derive our statistical theory in the form of a DDFT. Our starting point is the microscopic Smoluchowski equation for the interacting individual model microswimmers. 
Next, in Sec.~\ref{numerics}, details of a two-dimensional numerical implementation are listed together with the numerical results presented for a system under spherically symmetric confinement. 
In agreement with previous particle-based simulations \cite{nash2010run, hennes2014self} 
we observe a rotational symmetry breaking in certain parameter ranges, which can be identified as a ``hydrodynamic fluid pump''. An additional novel instability of this state is identified. Finally, we conclude in Sec.~\ref{conclusion}.

\section{Model}
\label{Model}

To derive our theory, we consider a dilute suspension of $N$ identical self-propelled microswimmers at low Reynolds number \cite{purcell1977life}. 
In particular, hydrodynamic interactions between these swimmers are to be included. The self-propulsion of a microswimmer is concatenated to self-induced fluid flows in the surrounding medium. This represents a major source of hydrodynamic interaction between different swimmers. To capture the effect, it is necessary to specify the geometry of the individual microswimmers, which sets the self-induced fluid flows. We proceed by first introducing a maximally reduced model microswimmer and then formulating the resulting interactions between pairs of such swimmers.

\subsection{Individual Microswimmer}

\begin{figure}
\includegraphics[width=7.3cm]{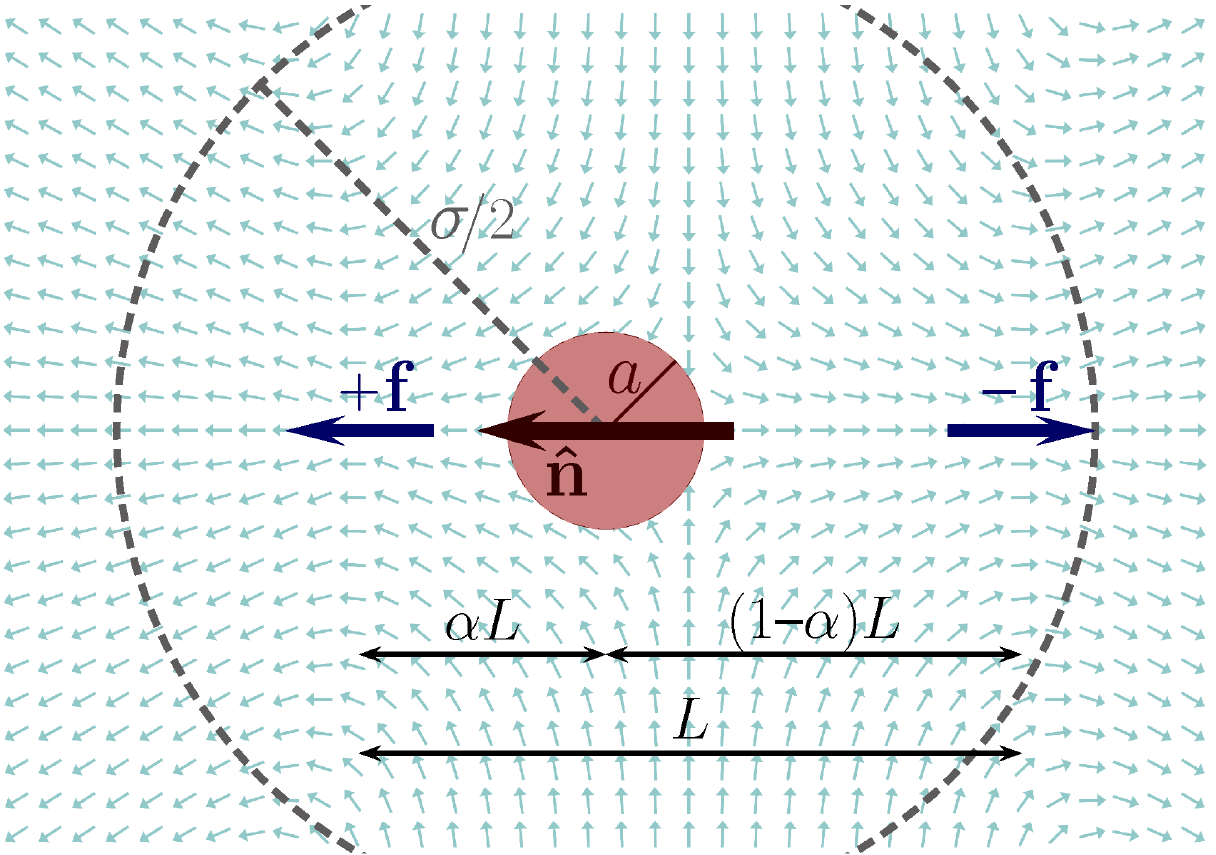}
\caption{
Individual model microswimmer. The spherical swimmer body of hydrodynamic radius $a$ is subjected to hydrodynamic drag. Two active point-like force centers exert active forces $+{\bf f}$ and $-{\bf f}$ onto the surrounding fluid. This results in a self-induced fluid flow indicated by small light arrows. $L$ is the distance between the two force centers. The whole set-up is axially symmetric with respect to the axis $\mathbf{\hat n}$. If the swimmer body is shifted along $\mathbf{\hat n}$ out of the geometric center, leading to distances $\alpha L$ and $(1-\alpha)L$ to the two force centers, it feels a net self-induced hydrodynamic drag. The microswimmer then self-propels. In the depicted state (pusher), fluid is pushed outward. Upon inversion of the two forces, fluid is pulled inward (puller). We consider soft isotropic steric interactions between the swimmer bodies of typical interaction range $\sigma$, implying an effective steric swimmer radius of $\sigma/2$. }
\label{fig_microswimmer}
\end{figure}

To keep the derivation and presentation of the theory in the following sections as simple as possible, we introduce a minimum model microswimmer as depicted in Fig.~\ref{fig_microswimmer}. 
Similar set-ups were mentioned in Refs. \cite{aditi2002hydrodynamic, hatwalne2004rheology, golestanian2008three, baskaran2009statistical}. 
Each microswimmer consists of a spherical body of hydrodynamic radius $a$. The swimmer body is subjected to hydrodynamic drag with respect to surrounding fluid flows. In this way, the swimmer can be convected by external flow fields. One way of self-convection is to generate a self-induced fluid flow. For this purpose, each microswimmer features two active force centers. They are located at a distance $L$ from each other on a symmetry axis that has orientation $\mathbf{\hat n}$ and runs through the center of the swimmer body. The two force centers exert two antiparallel forces $+\mathbf f$ and $-\mathbf f$, respectively, onto the surrounding fluid and set it into motion. Summing up the two forces, we find that the microswimmer exerts a vanishing net force onto the fluid. Moreover, since $\mathbf f\|{\mathbf{\hat n}}$, there is no net active torque \cite{fily2012cooperative}. The force centers are point-like and do not experience any hydrodynamic drag.

Self-propulsion is now achieved by shifting the swimmer body along $\mathbf{\hat n}$ out of the geometric center. We introduce a parameter $\alpha$ to quantify this shift, see Fig.~\ref{fig_microswimmer}. The distances between the body center and the force centers are now $\alpha L$ and $(1-\alpha)L$, respectively. We confine $\alpha$ to the interval $]0,0.5]$. For $\alpha=0.5$, the body is symmetrically located between the two force centers, and no net self-induced motion occurs. This geometry is called shaker \cite{hatwalne2004rheology,baskaran2009statistical}. For $\alpha\neq0.5$, the symmetry is broken. The swimmer body feels a net self-induced fluid flow due to the proximity to one of the two force centers. Due to the resulting self-induced hydrodynamic drag on the swimmer body, the swimmer self-propels. In the depicted state of outward oriented forces, the swimmer pushes the fluid outward and is called a pusher \cite{baskaran2009statistical}. Inverting the forces, the swimmer pulls fluid inward and is termed a puller \cite{baskaran2009statistical}.

\subsection {Hydrodynamic interactions}

We now consider an assembly of $N$ interacting identical self-propelled model microswimmers, suspended in a viscous, incompressible fluid at low Reynolds number \cite{purcell1977life}. The flow profile within the system then follows Stokes' equation \cite {happel2012low},
\begin{eqnarray}
{}-\eta\nabla^2\mathbf v(\mathbf r,t)+\nabla p(\mathbf r,t)=\sum_{i=1}^N\mathbf f_i(\mathbf r_i,\mathbf{\hat n}_i,t). 
\label{Stokes}
\end{eqnarray}
Here, $t$ denotes time and $\mathbf r$ any spatial position in the suspension, while, \rev{on the left-hand side,} $\mathbf v(\mathbf r,t)$ gives the corresponding fluid flow velocity field. $\eta$ is the viscosity of the fluid and $p(\mathbf r,t)$ is the pressure field. On the right-hand side, $\mathbf f_i$ denotes the total force density field exerted by the $i$th microswimmer onto the fluid. $\mathbf r_i$ and $\mathbf{\hat n}_i$ mark the current position and orientation of the $i$th swimmer at time $t$, respectively.

Obviously, on the one hand, each microswimmer contributes to the overall fluid flow in the system by the force density it exerts on the fluid. On the other hand, we have noted above that each swimmer is dragged along by the induced fluid flow. In this way, each swimmer can transport itself via active self-propulsion. Moreover, all swimmers hydrodynamically tear on each other via their induced flow fields. That is, they hydrodynamically interact, which influences their positions $\mathbf r_i$ and orientations $\mathbf{\hat n}_i$.

Progress can be made due to the linearity of Eq.~(\ref{Stokes}) and \rev{assuming} incompressibility of the fluid, i.e. $\nabla \cdot \mathbf v (\mathbf r,t)=0$. We denote by $\mathbf F_j$ and $\mathbf T_j$ the forces and torques, respectively, acting directly on the swimmer bodies ($j=1,...,N$), except for frictional forces and \rev{frictional} torques resulting from the surrounding fluid. 
The non-hydrodynamic body forces and torques may for example result from external potentials or steric interactions and will be specified below. From them, in the passive case, i.e.\ for $\mathbf f=\mathbf {0}$, the instantly resulting velocity  $\mathbf v_i$ and angular velocity $\bm{\omega}_i$ of the $i$th swimmer body follows as 
\begin{equation}
     \begin{bmatrix}
         \mathbf v_i \\
          \bm{\omega}_i
     \end{bmatrix}
     =
    \sum_{j=1}^N
    {\bf {M}}_{ij}\cdot\begin{bmatrix}
        \mathbf F_j \\
        \mathbf T_j
    \end{bmatrix}
    =
    \sum_{j=1}^N
    \begin{bmatrix}
    \bm{\mu}^{tt}_{ij} & \bm{\mu}^{tr}_{ij}\\
    \bm{\mu}^{rt}_{ij} & \bm{\mu}^{rr}_{ij}    
    \end{bmatrix}
    \cdot
    \begin{bmatrix}
        \mathbf F_j \\
        \mathbf T_j
    \end{bmatrix}.  
    \label{Stokes1}  
\end{equation} 
Here ${\bf M}_{ij}$ are the mobility matrices, the components of which ($\bm{\mu}^{tt}_{ij}$, $\bm{\mu}^{tr}_{ij}$, $\bm{\mu}^{rt}_{ij}$, $\bm{\mu}^{rr}_{ij}$) likewise form matrices.  
They describe hydrodynamic translation--translation, translation--rotation, rotation--translation, and rotation--rotation coupling, respectively.

This formalism is the same as for suspensions of passive colloidal particles \cite{dhont1996introduction,reichert2004hydrodynamic}. We consider stick boundary conditions for the fluid flow on the surfaces of the swimmer bodies. The microswimmers are assumed to be suspended in an infinite bulk fluid, where the fluid flow vanishes at infinitely remote distances. 
Then, there are several methods to determine the mobility matrices, e.g.\ the so-called method of reflections \cite{brenner1963stokes, dhont1996introduction} or the method of induced force multipoles \cite{mazur1982many}. 
In general, for $N$ interacting suspended particles, there is no exact analytical solution to the problem. Yet, the mobility matrices can be calculated in the form of a power series in $a/r_{ij}$. Here, $r_{ij}$ is the distance between the centers of the $i$th and $j$th swimmer body, i.e.\ 
$r_{ij}=|\mathbf r_{ij}|$ with $\mathbf r_{ij}=\mathbf r_j-\mathbf r_i$. 
The denser the suspension, the higher the orders in $a/r_{ij}$ that need to be taken into account for a reliable characterization. In the following, we confine ourselves to relatively dilute and semi-dilute systems, taking into account pairwise hydrodynamic interactions up to and including order $(a/r_{ij})^3$. In contrast to this, see for example Refs.~\cite{evans2011orientational,alarcon2013spontaneous,zottl2014hydrodynamics} for simulation approaches to dense suspensions of microswimmers.

To the order of $(a/r_{ij})^3$, hydrodynamic coupling is calculated in the following standard way. Since our system is overdamped, the forces $\mathbf F_j$ and torques $\mathbf T_j$ acting on the swimmer bodies are directly transmitted to the surrounding fluid. The fluid flow induced by each spherical swimmer body of hydrodynamic radius $a$ is calculated on the Rodne-Prager level \cite{dhont1996introduction}. At the position of the $i$th swimmer, the flow field induced by swimmer $j\neq i$ reads \cite{dhont1996introduction}
\begin{eqnarray}
\mathbf v(\mathbf r_i) &=&  \frac{1}{6\pi\eta a}\bigg(\frac{3a}{4r_{ij}}\left({\bf {1}}+\mathbf{\hat r}_{ij}\mathbf{\hat r}_{ij}\right) 
\nonumber\\
&&\qquad{}+\frac{a^3}{4r_{ij}^3}({\bf {1}}-3\mathbf{\hat r}_{ij}\mathbf{\hat r}_{ij})\bigg)\cdot{\mathbf F_j} \nonumber \\ 
&& {}+\frac{1}{8\pi\eta r_{ij}^3} \mathbf r_{ij} \times \mathbf T_j,
\label{Rodne-Prager}
\end{eqnarray}
where $\mathbf{1}$ is the unity matrix and $\mathbf{\hat r}_{ij}=\mathbf r_{ij}/r_{ij}$. The velocity $\mathbf v_i$ and angular velocity $\bm{\omega}_i$ resulting due to this flow field for the $i$th swimmer of hydrodynamic radius $a$ follows from Fax\'en's laws \cite{dhont1996introduction,navarro2010hydrodynamic}:  
\begin{eqnarray}
\mathbf v_i &=&  \left(1 +\frac{a^2}{6}\nabla_i^2\right)\mathbf v(\mathbf r_i), 
\label{faxen_v} \\
\bm{\omega}_i &=& \frac{1}{2}\nabla_i\times \mathbf v(\mathbf r_i).
\label{faxen_omega}
\end{eqnarray}
Due to the linearity of Stokes' equation, Eq.~(\ref{Stokes}), the overall velocities and angular velocities are obtained by superimposing the influence of all other swimmer bodies $j\neq i$. In addition to that, the direct effect of $\mathbf F_i$ and $\mathbf T_i$ on the motion of the $i$th swimmer is given by Stokes' drag formulae \cite{dhont1996introduction}
\begin{eqnarray}
\mathbf v_i &=& \frac{1}{6\pi\eta a}\mathbf F_i, \\
\bm{\omega}_i &=& \frac{1}{8\pi\eta a^3}\mathbf T_i. 
\end{eqnarray}

\rev{Combining} all these ingredients, the motion resulting for $\mathbf f=\mathbf 0$ can be conveniently summarized in the form of Eq.~(\ref{Stokes1}) by setting \cite{dhont1996introduction,reichert2004hydrodynamic}
\begin{eqnarray}
\bm{\mu} ^{tt}_{ii}&=&\mu^t{\bf 1},
\quad\bm{\mu}_{ii}^{rr}=\mu^r{\bf 1},
\quad\bm{\mu}_{ii}^{tr}=\bm{\mu}_{ii}^{rt}=\mathbf{0}\quad
\label{Stokes2}
\end{eqnarray} 
for entries $i=j$ (no summation over $i$ in these expressions) and
\begin{eqnarray}
\bm{\mu}_{ij}^{tt}&=&\mu^t\bigg(\frac{3a}{4r_{ij}}\Big({\bf {1}}+{{\mathbf{\hat r}_{ij}\mathbf{\hat r}_{ij}}}\Big)  \nonumber\\ 
&&{}+\frac{1}{2}\Big(\frac{a}{r_{ij}}\Big)^3\Big({\bf {1}}-3{{\mathbf{\hat r}_{ij}\mathbf{\hat r}_{ij}}}\Big)\bigg), 
\label{mu_tt} \\ 
\bm{\mu}_{ij}^{rr}&=&{}-\mu^r\frac{1}{2}\left(\frac{a}{r_{ij}}\right)^3\left({\bf {1}}-3{{\mathbf{\hat r}_{ij}\mathbf{\hat r}_{ij}}}\right), \\ 
\bm{\mu}_{ij}^{tr}&=&\bm{\mu}_{ij}^{rt}=\mu^r\left(\frac{a}{r_{ij}}\right)^3{ {\mathbf r_{ij}}}\times, 
\label{mu_tr}
\end{eqnarray} 
for entries $i\neq j$. Here, we have introduced the abbreviations
\begin{equation}\label{abbr}
\mu^t=\frac{1}{6\pi\eta a}, \qquad \mu^r=\frac{1}{8\pi\eta a^3}.
\end{equation}   
In this notation, the matrices $\bm{\mu}_{ij}^{tr}=\bm{\mu}_{ij}^{rt}$ in Eq.~(\ref{mu_tr}) represent operators with ``$\times$'' the vector product \cite{reichert2004hydrodynamic}.

So far, only the influence of the passive swimmer bodies has been included. We now take into account the active forces. Again, because of the linearity of Eq.~(\ref{Stokes}), their effect can simply be added to the swimmer velocities and angular velocities on the right-hand side of \rev{Eq.~(\ref{Stokes1})}.

The concept to include the influence of the active forces is the same as summarized above for the passive hydrodynamic interactions. There is only one difference. We consider the active force centers as point-like, and not of finite hydrodynamic radius. Moreover, they do not transmit torques to the fluid. Thus, instead of Eq.~(\ref{Rodne-Prager}), their induced flow fields are readily described on the Oseen level \cite{dhont1996introduction}. 
\rev{The} flow fields induced by the two force centers of the $j$th microswimmer at the position of the $i$th swimmer body read
\begin{eqnarray}
\mathbf v^+(\mathbf r_i) &=& \frac{1}{8\pi\eta r_{ij}^+}\left({\bf {1}}+\mathbf{\hat r}_{ij}^+\mathbf{\hat r}_{ij}^+\right)\cdot f{\mathbf{\hat n}_j},
\label{vplus}
\\
\mathbf v^-(\mathbf r_i) &=& {}-\frac{1}{8\pi\eta r_{ij}^-}\left({\bf {1}}+\mathbf{\hat r}_{ij}^-\mathbf{\hat r}_{ij}^-\right)\cdot f{\mathbf{\hat n}_j}. \qquad
\label{vminus}
\end{eqnarray}
These expressions are valid also for $i=j$, which leads to self-propulsion of a single isolated swimmer. We have defined 
\begin{eqnarray}
\mathbf r_{ij}^+ &=& \mathbf r_{ij}+\alpha L \mathbf{\hat n}_j, \label{defplus}
\\
\mathbf r_{ij}^- &=& \mathbf r_{ij}-(1-\alpha) L \mathbf{\hat n}_j \label{defminus}
\end{eqnarray}
to refer to the distance vectors between the active force centers of the $j$th swimmer and the center of the $i$th swimmer body. Moreover, we have parameterized 
\begin{equation}
\mathbf f_j = f\mathbf{\hat n}_j
\end{equation}
so that the sign of $f$ now determines the character of the swimmer (pusher or puller).

In analogy to the passive case, the velocities and angular velocities of the swimmer bodies of finite hydrodynamic radius $a$ that result from the active flow fields Eqs.~(\ref{vplus}) and (\ref{vminus}) are calculated from Fax\'en's laws, Eqs.~(\ref{faxen_v}) and (\ref{faxen_omega}). 
The result can be written using mobility matrices 
\begin{eqnarray}
\bm{\mu}_{ij}^{tt\pm} &=&
\frac{1}{8\pi\eta r_{ij}^{\pm}}\left({\bf {1}}+\mathbf{\hat r}_{ij}^{\pm}\mathbf{\hat r}_{ij}^{\pm}\right) \nonumber \\
&&{}+\frac{a^2}{24\pi\eta {r_{ij}^{\pm}}^3}\left({\bf {1}}-3\mathbf{\hat r}^{\pm}_{ij}\mathbf{\hat r}^{\pm}_{ij}\right), 
\label{mu_tt_pm}\\
\bm{\mu}_{ij}^{rt\pm} &=& \frac{1}{8\pi\eta {r_{ij}^{\pm}}^3}\mathbf r_{ij}^{\pm}\times.
\label{mu_rt_pm}
\end{eqnarray} 
Within this framework, the corresponding active forces on the right-hand side of Eq.~(\ref{Stokes1}) have to be inserted as $\pm f\mathbf{\hat n}_j$. Since there are no active torques, we may set $\bm{\mu}_{ij}^{tr\pm} = \bm{\mu}_{ij}^{rr\pm} = \bm{0}$. Altogether, \rev{passive and active hydrodynamic interactions}, including the self-propulsion mechanism, \rev{are now formulated up to third order in $a/r_{ij}$}.

\subsection{Body forces and torques}

We now specify the non-hydrodynamic forces $\mathbf F_j$ and torques $\mathbf T_j$ acting directly on the swimmer bodies. In our case, these forces can be written as 
\begin{equation}
\mathbf F_j =-\nabla_j U -{\nabla_j\ln P}.
\label{force}
\end{equation}
Here, $\nabla_j$ denotes the partial derivative $\partial/\partial\mathbf r_j$. Throughout this work, we measure energies in units of $k_BT$ with $k_B$ the Boltzmann constant and $T$ the temperature of the fluid. Variations in temperature due to the non-equilibrium nature of our system are ignored. 
In Eq.~(\ref{force}), the first contribution results from a potential
\begin{equation}
U\left(\mathbf r^N\right)=\frac{1}{2}\sum_{\substack{k,l=1\\k\neq l}}^Nu(\mathbf r_k,\mathbf r_l)+\sum_{l=1}^N u_{ext}(\mathbf r_l),
\label{eq_U}
\end{equation}
where we use the abbreviation $\mathbf r^N=\{\mathbf r_1,\mathbf r_2,...\mathbf r_N\}$. Accordingly, we will abbreviate $\mathbf{\hat n}^N=\{\mathbf{\hat n}_1,\mathbf{\hat n}_2,...\mathbf{\hat n}_N\}$ below. For simplicity and as a first step, we confine ourselves to soft pairwise steric interactions of the form
\begin{equation}
u(\mathbf r_k,\mathbf r_l)=\epsilon_0\exp\left(-\frac{r_{kl}^4}{\sigma^4}\right). 
\label{eqGEM4}
\end{equation}
$\epsilon_0$ sets the strength of this potential and $\sigma$ an effective interaction range, see Fig.~\ref{fig_microswimmer}. Such soft interaction potentials are frequently employed to describe effective interactions in soft matter systems, e.g.\ between polymers, star-polymers, dendrimers, and other macromolecules in solution \cite{mladek2006formation}. One task for the future is to clarify more precisely the nature of the effective steric interactions between individual microswimmers, for instance for self-propelling microorganisms featuring agitated cilia and flagella \cite{wensink2014controlling}. 
We prefer the so-called GEM-4 potential in Eq.~(\ref{eqGEM4}) to a simple Gaussian interaction, because it can describe both liquid and solid phases within mean-field approximation, in contrast to the Gaussian potential \cite{archer2014solidification}. The phase behavior depends on the parameter $\epsilon_0$ as well as on the average density of the suspended particles. Here, we fix the parameters such that our system remains in the liquid phase. Moreover, the density is adjusted to avoid overlap of the swimmers. Properties of crystallized systems may be investigated in a later study. 

In addition to that, we consider the microswimmers to be confined to a rotationally symmetric external trapping potential. It constitutes the second contribution on the right-hand side of Eq.~(\ref{eq_U}) and reads 
\begin{equation}
u_{ext}(\mathbf r_l)=k|\mathbf r_{l}|^4.
\end{equation} 
$k$ sets the strength of the trap. We choose the quartic potential instead of a more common harmonic trap due to its lower gradient at smaller radii. Overlap of individual swimmers is reduced in this way.

The quantity $P\equiv P(\mathbf r^N,\mathbf{\hat n}^N,t)$ in Eq.~(\ref{force}) denotes the probability distribution to find the $N$ microswimmers at time $t$ at positions $\mathbf r^N$ with orientations \rev{$\mathbf{\hat n}^N$}. 
Via the contribution involving $\ln P$, we consistently include entropic forces into our statistical characterization \cite{doi1988theory}. This term represents the effect of thermal forces acting on each swimmer as a result of thermal fluctuations.

Due to the spherical shape of the swimmer bodies, and for simplicity, we assume in the present work that non-hydrodynamic torques acting on the swimmer bodies solely result from thermal fluctuations. They can be included into our statistical formalism by setting \cite{doi1988theory}
\begin{eqnarray}
{\mathbf T_j} = -\mathbf{\hat n}_j\times\nabla_{\mathbf{\hat n}_j}\ln P.
\label{torque}
\end{eqnarray}          
Further contributions to the torques, e.g.\ resulting from steric alignment interactions between different swimmers, may be considered in future studies.

\section{Derivation of the DDFT for microswimmers}
\label{Derivation}

Our starting point to derive the dynamical density functional theory (DDFT) for active microswimmers including hydrodynamic interactions is the microscopic Smoluchowski equation \cite{doi1988theory}  for $N$ identical interacting swimmers. This continuity equation for the time evolution of the probability distribution $P(\mathbf r^N, \mathbf{\hat n}^N, t)$ reads
\begin{equation}
\frac{\partial P}{\partial t}= -\sum_{i=1}^N\Big(\nabla_i\cdot\left(\mathbf v_i P\right)+\left(\mathbf{\hat n}_i\times \nabla_{\mathbf{\hat n}_i}\right)\cdot\left(\bm{\omega}_i P\right)\Big).
\label{Smoluchowski}
\end{equation} 
On the basis of Sec.~\ref{Model}, we insert
\begin{eqnarray}
\mathbf v_i &=& \sum_{j=1}^N\left(\bm{\mu}_{ij}^{tt}\cdot\mathbf F_j+\bm{\mu}_{ij}^{tr}\, \mathbf T_j + \bm{\Lambda}_{ij}^{tt}\cdot\mathbf{\hat n}_jf\right), \quad
\label{eq_vi}\\
\bm{\omega}_i &=& \sum_{j=1}^N\left(\bm{\mu}_{ij}^{rt}\, \mathbf F_j+\bm{\mu}_{ij}^{rr}\cdot\mathbf T_j + \bm{\Lambda}^{rt}_{ij}\, \mathbf{\hat n}_jf\right), 
\label{eq_omegai}
\end{eqnarray}
where we have introduced the abbreviations 
\begin{eqnarray}
\bm{\Lambda}_{ij}^{tt} &=& \bm{\mu}_{ij}^{tt+}-\bm{\mu}_{ij}^{tt-},
\label{Lambda_tt} \\
\bm{\Lambda}_{ij}^{rt} &=& \bm{\mu}_{ij}^{rt+}-\bm{\mu}_{ij}^{rt-}. 
\label{Lambda_rt}
\end{eqnarray}
Thus the hydrodynamic interactions enter via the configuration-dependent expressions for $\mathbf v_i$ and $\bm{\omega}_i$. 
For a single, isolated microswimmer, i.e.\ for $N=1$, the self-propulsion velocity becomes  $\mathbf v_1 = \bm{\Lambda}_{11}^{tt}\cdot\mathbf{\hat n}_1f$, which is directed along the swimmer axis and vanishes in the case of a shaker, where $\alpha=0.5$.

Our scope is to derive from Eq.~(\ref{Smoluchowski}) a dynamic equation for the swimmer density $\rho^{(1)}(\mathbf r,\mathbf{\hat n},t)$. In general, the $n$-swimmer density $\rho^{(n)}(\mathbf r^n,\mathbf{\hat n}^n,t)$ for $n\leq N$ is obtained from the probability distribution $P(\mathbf r^N, \mathbf{\hat n}^N,t)$ by integrating out the degrees of freedom of $N-n$ swimmers,
\begin{widetext} 
\begin{equation}
\rho^{(n)}(\mathbf r^n,\mathbf{\hat n}^n,t) = \frac{N!}{(N-n)!}\int \mathrm{d}\mathbf r_{n+1}\int \mathrm{d}\mathbf{\hat n}_{n+1}\dots 
\int \mathrm{d}\mathbf r_{N} \int \mathrm{d}\mathbf{\hat n}_{N}\;P(\mathbf r^N, \mathbf{\hat n}^N,t).
\label{nbody}
\end{equation}
Accordingly, we obtain a dynamic equation for $\rho^{(1)}(\mathbf r,\mathbf{\hat n},t)$ by integrating out from Eq.~(\ref{Smoluchowski}) the degrees of freedom of $N-1$ swimmers. This leads us to
\begin{equation}
\label{BBGKY1}
\frac{\partial\rho^{(1)}(\mathbf r,\mathbf{\hat n},t)}{\partial t} = -\nabla_{\mathbf r}\cdot(\bm{\mathcal{ J}\!}_1+\bm{\mathcal{ J}\!}_2+\bm{\mathcal{ J}\!}_3) 
-(\mathbf{\hat n} \times \nabla_{\mathbf{\hat n}})\cdot(\bm{\mathcal{ J}\!}_4+\bm{\mathcal{ J}\!}_5+\bm{\mathcal{ J}\!}_6),
\end{equation}
with the abbreviations
\begin{eqnarray}
\bm{\mathcal{ J}\!}_1
&=&
{}-\mu^t\left(\nabla_{\mathbf r}\,\rho^{(1)}(\mathbf r,\mathbf{\hat n},t)+\rho^{(1)}(\mathbf r,\mathbf{\hat n},t)\nabla_{\mathbf r}\,u_{ext}(\mathbf r) 
+\int \mathrm{d}\mathbf r' \mathrm{d}\mathbf{\hat n}' \rho^{(2)}(\mathbf r, \mathbf r^{\prime},\mathbf{\hat n}, \mathbf{\hat n}^{\prime}, t)\nabla_{\mathbf r}u(\mathbf r, \mathbf r')\right) 
\nonumber\\ &&
{}-\int \mathrm{d}\mathbf r' \mathrm{d}\mathbf{\hat n}' \,{\bm{\mu}^{tt}_{\mathbf r, \mathbf r'}}\cdot\bigg(\nabla_{\mathbf r'}\rho^{(2)}(\mathbf r, \mathbf r', \mathbf{\hat n},\mathbf{\hat n}',t)
+\rho^{(2)}(\mathbf r,\mathbf r',\mathbf{\hat n}, \mathbf{\hat n}', t)\nabla_{\mathbf r'}u_{ext}(\mathbf r')\nonumber\\ &&
\qquad{}+\rho^{(2)}(\mathbf r, \mathbf r',\mathbf{\hat n}, \mathbf{\hat n}', t)\nabla_{\mathbf r'}u({\mathbf r,\mathbf r')} 
+\int \mathrm{d}\mathbf r'' \mathrm{d}\mathbf{\hat n}'' \rho^{(3)}(\mathbf r,{\mathbf {r}}', {\mathbf {r}}'',\mathbf{\hat n}, \mathbf{\hat n}', \mathbf{\hat n}'', t)\nabla_{\mathbf r'}u(\mathbf r',\mathbf r'')\bigg),
\label{eqJ1} \\ 
\bm{\mathcal{ J}\!}_2
&=&
{}-\int \mathrm{d}\mathbf r' \mathrm{d}\mathbf{\hat n}' \,\bm{\mu}_{\mathbf r, \mathbf r'}^{tr} (\mathbf{\hat n}'\times\nabla_{\mathbf{\hat n}'})\rho^{(2)}(\mathbf r, \mathbf r',\mathbf{\hat n},\mathbf{\hat n}', t),
\label{eqJ2} \\ 
\bm{\mathcal{ J}\!}_3
&=&
f\left({\bm{\Lambda}^{tt}_{\mathbf r, \mathbf r}}\cdot\mathbf{\hat n} \rho^{(1)}(\mathbf r,\mathbf{\hat n}, t)   
+\int \mathrm{d}\mathbf r' \mathrm{d}\mathbf{\hat n}' \,{\bm{\Lambda}^{tt}_{\mathbf r, \mathbf r'}}\cdot\mathbf{\hat n}'\rho^{(2)}(\mathbf r, \mathbf r',\mathbf{\hat n},\mathbf{\hat n}', t)\right),
\label{eqJ3} \\ 
\bm{\mathcal{ J}\!}_4
&=&
{}-\int \mathrm{d}\mathbf r' \mathrm{d}\mathbf{\hat n}' {\bm{\mu}^{rt}_{\mathbf r, \mathbf r'}} \bigg(\nabla_{\mathbf r'}\rho^{(2)}(\mathbf r, \mathbf r', \mathbf{\hat n},\mathbf{\hat n}',t) 
+\rho^{(2)}(\mathbf r,\mathbf r',\mathbf{\hat n}, \mathbf{\hat n}', t)\nabla_{\mathbf r'}u_{ext}(\mathbf r')
\nonumber \\ &&
\qquad{}+\rho^{(2)}(\mathbf r, \mathbf r',\mathbf{\hat n}, \mathbf{\hat n}', t)\nabla_{\mathbf r'}u({\mathbf r,\mathbf r')} + \int \mathrm{d}\mathbf r'' \mathrm{d}\mathbf{\hat n}'' \rho^{(3)}(\mathbf r,{\mathbf {r}}', {\mathbf {r}}'',\mathbf{\hat n}, \mathbf{\hat n}', \mathbf{\hat n}'', t)\nabla_{\mathbf r'}u(\mathbf r',\mathbf r'')\bigg),
\label{eqJ4} \\ 
\bm{\mathcal{ J}\!}_5
&=&
{}-\mu^r\mathbf{\hat n}\times \nabla_{\mathbf{\hat n}} \rho^{(1)}(\mathbf r,\mathbf{\hat n}, t)-
\int \mathrm{d}\mathbf r'\mathrm{d}\mathbf{\hat n}' \,\bm{\mu}^{rr}_{\mathbf r, \mathbf r'}\cdot(\mathbf{\hat n}'\times\nabla_{\mathbf n'})\rho^{(2)}(\mathbf r,\mathbf r',\mathbf{\hat n},\mathbf{\hat n}',t),
\label{eqJ5} \\ 
\bm{\mathcal{ J}\!}_6
&=&
f\int \mathrm{d}\mathbf r' \mathrm{d}\mathbf{\hat n}' \,{\bm{\Lambda}^{rt}_{\mathbf r, \mathbf r'}}\, \mathbf{\hat n}' \rho^{(2)}(\mathbf r, \mathbf r',\mathbf{\hat n},\mathbf{\hat n}', t).
\label{eqJ6}
\end{eqnarray}

\end{widetext}

Eq.~(\ref{BBGKY1}) represents the dynamic equation for our searched-for quantity $\rho^{(1)}$. However, as a consequence of the inter-swimmer interactions within our system, the equation contains the unknown two- and three-swimmer densities $\rho^{(2)}$ and $\rho^{(3)}$. Dynamic equations for these higher-$n$ swimmer densities can likewise be derived from Eq.~(\ref{Smoluchowski}) by integrating out the degrees of freedom of $N-n$ swimmers. Yet, this only shifts the problem to higher $n$. It is found that the dynamic equation for $\rho^{(n)}$ contains $\rho^{(n+1)}$ and $\rho^{(n+2)}$ for $1\leq n\leq N-2$. Therefore, a reliable closure scheme is needed to cut this hierarchy of coupled dynamic partial differential equations, typically referred to as the Bogoliubov-Born-Green-Kirkwood-Yvon (BBGKY) hierarchy \cite{hansen1990theory}. 
DDFT provides such closure relations. In the following, we employ this approach to break the hierarchy already at $n=1$. Thus we derive a decoupled dynamic equation for $\rho^{(1)}(\mathbf r, \mathbf{\hat n}, t)$.

DDFT uses as an input the concepts from equilibrium density functional theory (DFT) \cite{hansen1990theory,singh1991density,evans1992density,marconi1999dynamic, marconi2000dynamic, archer2004dynamical,evans2010density,lowen2010density}. Most importantly, DFT implies that a certain observed equilibrium density $\rho^{(1)}_{eq}(\mathbf r,\mathbf{\hat n})$ can only result from one unique external potential $\Phi_{ext}(\mathbf r,\mathbf{\hat n})$ acting on the system. As a consequence, $\Phi_{ext}(\mathbf r,\mathbf{\hat n})$ is set by an observed $\rho^{(1)}_{eq}(\mathbf r,\mathbf{\hat n})$ and, moreover, the grand canonical potential ${\Omega}$ and the free energy $\mathcal{F}$ can be expressed as functionals of $\rho^{(1)}(\mathbf r,\mathbf{\hat n})$. In our case, we may write
\begin{equation}
\label{Omega}
\Omega\left[\rho^{(1)}\right]
=
\mathcal{F}_{id}\left[\rho^{(1)}\right] +\mathcal{F}_{exc}\left[\rho^{(1)}\right] 
+\mathcal{F}_{ext}\left[\rho^{(1)}\right].
\end{equation}
Here
\begin{eqnarray}
\mathcal{F}_{id}\left[\rho^{(1)}\right]
&=&
\int \mathrm{d}\mathbf r\, \mathrm{d}\mathbf{\hat n}\,\rho^{(1)}(\mathbf r,\mathbf{\hat n})
\nonumber\\
&&\quad{}\times\left(\ln\left(\lambda^3\rho^{(1)}(\mathbf r,\mathbf{\hat n})\right)-1\right)\qquad
\end{eqnarray}
is the entropic contribution for an ideal gas of non-interacting particles with $\lambda$ the thermal de Broglie wave length \cite{wittkowski2010derivation}. We recall that energies are measured in units of $k_BT$ throughout this work. 
Next, the excess free energy $\mathcal{F}_{exc}$ contains all particle interactions, i.e.\ contributions beyond the limit of an ideal gas. $\mathcal{F}_{exc}$ is generally not known analytically and must be approximated. The third term reads
\begin{equation}
\mathcal{F}_{ext}\left[\rho^{(1)}\right] = \int \mathrm{d}\mathbf r\, \mathrm{d}\mathbf{\hat n}\,\Phi_{ext}(\mathbf r,\mathbf{\hat n})\rho^{(1)}(\mathbf r,\mathbf{\hat n}),  
\end{equation}
where here we have included the effect of a chemical potential into $\Phi_{ext}(\mathbf r,\mathbf{\hat n})$. In this form, DFT reduces to a variational problem to determine the equilibrium density,
\begin{equation}
\left.\frac{\delta\Omega}{\delta \rho^{(1)}(\mathbf r,\mathbf{\hat n})}\right|_{\rho^{(1)}(\mathbf r,\mathbf{\hat n})=\rho^{(1)}_{eq}(\mathbf r,\mathbf{\hat n})} = \;0. 
\end{equation}
Inserting Eq.~(\ref{Omega}) leads to
\begin{eqnarray}
\lefteqn{\ln\left(\lambda^3\rho^{(1)}_{eq}(\mathbf r,\mathbf{\hat n})\right)+\Phi_{ext}(\mathbf r,\mathbf{\hat n})\,=}\nonumber \\
&&\qquad\qquad{}-\left.\frac{\delta {\mathcal F}_{exc}}{\delta \rho^{(1)}(\mathbf r,\mathbf{\hat n})}\right|_{\rho^{(1)}(\mathbf r,\mathbf{\hat n})=\rho^{(1)}_{eq}(\mathbf r,\mathbf{\hat n})}. 
\label{rhoeq_equilibrium_prev}
\end{eqnarray}

The central approximation of DDFT is to transfer equilibrium relations to the non-equilibrium case. For this purpose, at each time $t$ and for the corresponding $\rho^{(1)}(\mathbf r,\mathbf{\hat n},t)$, one assumes an instantaneous external potential $\Phi_{ext}(\mathbf r,\mathbf{\hat n},t)$ that satisfies the above relations. In particular, we assume that Eq.~(\ref{rhoeq_equilibrium_prev}) 
still holds with $\rho^{(1)}_{eq}(\mathbf r,\mathbf{\hat n})$ and $\Phi_{ext}(\mathbf r,\mathbf{\hat n})$ replaced by $\rho^{(1)}(\mathbf r,\mathbf{\hat n},t)$ and $\Phi_{ext}(\mathbf r,\mathbf{\hat n},t)$, respectively, i.e.\ 
\begin{equation}
\ln\left(\lambda^3\rho^{(1)}(\mathbf r,\mathbf{\hat n},t)\right)+\Phi_{ext}(\mathbf r,\mathbf{\hat n},t)=
{}-\frac{\delta {\mathcal F}_{exc}}{\delta \rho^{(1)}(\mathbf r,\mathbf{\hat n},t)}. 
\label{rhoeq_prev}
\end{equation}

In combination with that, to close our dynamic equation for $\rho^{(1)}(\mathbf r,\mathbf{\hat n},t)$, we use relations that would follow from Eqs.~(\ref{eqJ1})--(\ref{eqJ6}) in static equilibrium. In this case, $f=0$ and $\bm{\mathcal{ J}\!}_3=\bm{\mathcal{ J}\!}_6=\mathbf 0$. Moreover, our interaction potentials and the external potential $u_{ext}$ do not depend on the swimmer orientations. Then, in equilibrium, it follows that $\mathbf{\hat n}\times\nabla_{\mathbf{\hat n}}\rho^{(n)}=\mathbf 0$ for all $n$ and therefore $\bm{\mathcal{ J}\!}_2=\bm{\mathcal{ J}\!}_5=\mathbf 0$. The remaining translational and rotational currents $\bm{\mathcal{ J}\!}_1$ and $\bm{\mathcal{ J}\!}_4$ must vanish independently of each other in static equilibrium. From these conditions, and replacing in the resulting expressions $u_{ext}(\mathbf r)$ by $\Phi_{ext}(\mathbf r,\mathbf{\hat n},t)$, which manifests the central DDFT approximation, we obtain
\begin{widetext}
\begin{equation}
\mathbf {0} = \nabla_{\mathbf r}\rho^{(1)}(\mathbf r,\mathbf{\hat n},t)+\rho^{(1)}(\mathbf r,\mathbf{\hat n},t)\nabla_{\mathbf r}\Phi_{ext}(\mathbf r,\mathbf{\hat n},t)
+\int \mathrm{d}\mathbf r'\mathrm{d}\mathbf{\hat n}' \rho^{(2)}(\mathbf r, \mathbf r',\mathbf{\hat n},\mathbf{\hat n}',t)\nabla_{\mathbf r}u(\mathbf r, \mathbf r')
\label{YYBG1}                   
\end{equation}
and
\begin{eqnarray}
\mathbf {0} 
&=&
\nabla_{\mathbf r'}\rho^{(2)}(\mathbf r, \mathbf r',\mathbf{\hat n},\mathbf{\hat n}',t)
+\rho^{(2)}(\mathbf r,\mathbf r',\mathbf{\hat n},\mathbf{\hat n}',t)\nabla_{\mathbf r'}\big(\Phi_{ext}(\mathbf r',\mathbf{\hat n}',t)+u(\mathbf r,\mathbf r')\big) 
\nonumber\\
&&{}+\int \mathrm{d}\mathbf r'' \mathrm{d}\mathbf{\hat n}''\rho^{(3)}(\mathbf r, \mathbf r',\mathbf r'',\mathbf{\hat n},\mathbf{\hat n}',\mathbf{\hat n}'',t)\nabla_{\mathbf r'}u(\mathbf r',\mathbf r'').  
\label{YYBG2}                  
\end{eqnarray}
Here, Eq.~(\ref{YYBG2}) was used to eliminate a major part in Eq.~(\ref{YYBG1}) that followed from the expression for $\bm{\mathcal{ J}\!}_1$. In fact, Eqs.~(\ref{YYBG1}) and (\ref{YYBG2}) are the first two members of a series of hierarchical relations, the so-called Yvon-Born-Green (YBG) relations, that can be derived in static equilibrium \cite{hansen1990theory}.

Now, inserting Eq.~(\ref{rhoeq_prev}) 
into Eqs.~(\ref{YYBG1}) and (\ref{YYBG2}) to eliminate the unknown potential $\Phi_{ext}(\mathbf r,\mathbf{\hat n},t)$, we find
\begin{equation}
\int \mathrm{d}\mathbf r' \mathrm{d}\mathbf{\hat n}' \rho^{(2)}(\mathbf r,\mathbf r',\mathbf{\hat n},\mathbf{\hat n}',t)\nabla_{\mathbf r'}u(\mathbf r,\mathbf r')
=
\rho^{(1)}(\mathbf r,\mathbf{\hat n},t)\nabla_{\mathbf r}\frac{\delta {\mathcal F}_{exc}}{\delta\rho^{(1)}(\mathbf r,\mathbf{\hat n},t)}
\label{YBG1}
\end{equation} 
and
\begin{eqnarray}
\lefteqn{
\nabla_{\mathbf r'}\rho^{(2)}(\mathbf r,\mathbf r',\mathbf{\hat n},\mathbf{\hat n}',t)
+\rho^{(2)}(\mathbf r,\mathbf r',\mathbf{\hat n},\mathbf{\hat n}',t)\nabla_{\mathbf r'}u(\mathbf r,\mathbf r')
+\int \mathrm{d}\mathbf r'' \mathrm{d}\mathbf{\hat n}''\rho^{(3)}(\mathbf r,\mathbf r',\mathbf r'',\mathbf{\hat n},\mathbf{\hat n}',\mathbf{\hat n}'',t)\nabla_{\mathbf r'}u(\mathbf r',\mathbf r'') \;=} \nonumber\\ 
&&\rho^{(2)}(\mathbf r,\mathbf r',\mathbf{\hat n},\mathbf{\hat n}',t)\left(\nabla_{\mathbf r'}\ln\left(\lambda^3\rho^{(1)}(\mathbf r',\mathbf{\hat n}',t)\right)
+\nabla_{\mathbf r'}\frac{\delta {\mathcal F}_{exc}}{\delta \rho^{(1)}(\mathbf r',\mathbf{\hat n}',t)}\right). \qquad\qquad\qquad\qquad\qquad\qquad\qquad
\label{YBG2}
\end{eqnarray} 
As a major benefit of this procedure, the three-swimmer density $\rho^{(3)}$ can be eliminated from the currents in Eqs.~(\ref{eqJ1})--(\ref{eqJ6}) by inserting Eq.~(\ref{YBG2}). Moreover, one occurrence of $\rho^{(2)}$ is eliminated using Eq.~(\ref{YBG1}). 
The currents then reduce to
\begin{eqnarray}
\label{DDFT_J1}
\bm{\mathcal{ J}\!}_1
&=&
{}-\mu^t\left(\nabla_{\mathbf r}\rho^{(1)}(\mathbf r,\mathbf{\hat n},t)+\rho^{(1)}(\mathbf r,\mathbf{\hat n},t)\nabla_{\mathbf r}u_{ext}(\mathbf r)
+\rho^{(1)}(\mathbf r,\mathbf{\hat n},t)\nabla_{\mathbf r}\frac{\delta {\mathcal F}_{exc}}{\delta\rho^{(1)}(\mathbf r,\mathbf{\hat n},t)}\right)
\nonumber\\&&
{}-\int \mathrm{d}\mathbf r' \mathrm{d}\mathbf{\hat n}' { \bm{\mu}^{tt}_{\mathbf r,\mathbf r'}}\cdot\left(\rho^{(2)}(\mathbf r,\mathbf r',\mathbf{\hat n},\mathbf{\hat n}',t)
\left(\nabla_{\mathbf r'}\ln\left(\lambda^3\rho^{(1)}(\mathbf r',\mathbf{\hat n}',t)\right) + \nabla_{\mathbf r'}u_{ext}(\mathbf r')
+\nabla_{\mathbf r'}\frac{\delta {\mathcal F}_{exc}}{\delta\rho^{(1)}(\mathbf r',\mathbf{\hat n}',t)}\right)\right),\\[.2cm]
\bm{\mathcal{ J}\!}_2
&=&
{}-\int \mathrm{d}\mathbf r' \mathrm{d}\mathbf{\hat n}' \,\bm{\mu}_{\mathbf r,\mathbf r'}^{tr}\,(\mathbf{\hat n}'\times\nabla_{\mathbf{\hat n}'})\rho^{(2)}(\mathbf r,\mathbf r',\mathbf{\hat n},\mathbf{\hat n}',t), \label{DDFT_J2}
\\[.2cm]
\bm{\mathcal{ J}\!}_3
&=&
f\left({\bm{\Lambda}^{tt}_{\mathbf r,\mathbf r}}\cdot\mathbf{\hat n}\rho^{(1)}(\mathbf r,\mathbf{\hat n},t)
+\int \mathrm{d}\mathbf {r}'\mathrm{d}\mathbf{\hat n}'{\bm{\Lambda}^{tt}_{\mathbf r,\mathbf r'}}\cdot\mathbf{\hat n}'\rho^{(2)}(\mathbf r,\mathbf r',\mathbf{\hat n},\mathbf{\hat n}',t)\right), \label{DDFT_J3}\\[.2cm] 
\bm{\mathcal{ J}\!}_4
&=&
{}-\int \mathrm{d}\mathbf r' \mathrm{d}\mathbf{\hat n}' { \bm{\mu}^{rt}_{\mathbf r,\mathbf r'}}\,\left(\rho^{(2)}(\mathbf r,\mathbf r',\mathbf{\hat n},\mathbf{\hat n}',t) 
\left(\nabla_{\mathbf r'}\ln\left(\lambda^3\rho^{(1)}(\mathbf r',\mathbf{\hat n}',t)\right) 
+\nabla_{\mathbf r'}u_{ext}(\mathbf r')+\nabla_{\mathbf r'}\frac{\delta {\mathcal F}_{exc}}{\delta\rho^{(1)}(\mathbf r',\mathbf{\hat n}',t)}\right)\right), \label{DDFT_J4}\\[.2cm]
\bm{\mathcal{ J}\!}_5
&=&
{}-\mu^r\mathbf{\hat n} \times \nabla_{\mathbf{\hat n}} \rho^{(1)}(\mathbf r,\mathbf{\hat n},t)
-\int \mathrm{d}\mathbf r'\mathrm{d}\mathbf{\hat n}' \,\bm{\mu}^{rr}_{\mathbf r,\mathbf r'}\cdot(\mathbf{\hat n}'\times\nabla_{\mathbf n'})\rho^{(2)}(\mathbf r,\mathbf r',\mathbf{\hat n},\mathbf{\hat n}',t), \label{DDFT_J5}\\[.2cm] 
\bm{\mathcal{ J}\!}_6&=&f\int \mathrm{d}\mathbf r' \mathrm{d}\mathbf{\hat n}' \,{\bm{\Lambda}^{rt}_{\mathbf r,\mathbf r'}}\,\mathbf{\hat n}'\rho^{(2)}(\mathbf r, \mathbf r',\mathbf{\hat n},\mathbf{\hat n}', t).
\label{DDFT_J6}
\end{eqnarray}

\end{widetext}

In effect, we have replaced $\rho^{(3)}(\mathbf r, \mathbf r', \mathbf r'', \mathbf{\hat n}, \mathbf{\hat n}', \mathbf{\hat n}'', t)$ and one instance of $\rho^{(2)}(\mathbf r, \mathbf r', \mathbf{\hat n}, \mathbf{\hat n}', t)$ by their equilibrium expressions that would apply, if the equilibrium one-swimmer density were given by $\rho^{(1)}(\mathbf r, \mathbf{\hat n},t)$. This procedure works best when $\rho^{(3)}$ and $\rho^{(2)}$ relax significantly quicker than $\rho^{(1)}$. It is therefore referred to as adiabatic elimination \cite{risken1984fokker}. In our case, the overdamped nature of the microswimmer dynamics supports this procedure.

Finally, we need to express ${\mathcal F}_{exc}$ and $\rho^{(2)}$ as functionals of $\rho^{(1)}$ to close the dynamical equation for $\rho^{(1)}(\mathbf r,\mathbf{\hat n},t)$. For moderate interaction strengths $\epsilon_0\lesssim 1$ in our soft GEM-4 interaction potential Eq.~(\ref{eqGEM4}), the classical mean-field approximation provides a reasonable and simple closure scheme \cite{archer2014solidification}. It is given by
\begin{equation} 
{\mathcal F}_{exc}=\frac{1}{2}\int \mathrm{d}\mathbf{r}\,\mathrm{d}\mathbf{r}'\mathrm{d}\mathbf{\hat n} \,\mathrm{d}\mathbf{\hat n}'\rho^{(2)}(\mathbf r,\mathbf r',\mathbf{\hat n},\mathbf{\hat n}',t)u(\mathbf r,\mathbf r')
\label{Fexc}
\end{equation} 
for the excess free energy and the approximation
\begin{equation} 
\rho^{(2)}(\mathbf r,\mathbf r',\mathbf{\hat n},\mathbf{\hat n}',t)=\rho^{(1)}(\mathbf r,\mathbf{\hat n},t)\,\rho^{(1)}(\mathbf r',\mathbf{\hat n}',t) 
\label{rho2}
\end{equation} 
for the two-swimmer density.

Overall, Eqs.~(\ref{BBGKY1}) and (\ref{DDFT_J1})--(\ref{DDFT_J6}) together with Eqs.~(\ref{Fexc}) and (\ref{rho2}) complete our derivation of a DDFT for dilute to semi-dilute suspensions of active microswimmers. We included hydrodynamic and soft steric interactions. Inserting the mobility tensors listed in \rev{Eqs.~(\ref{Stokes2})--(\ref{abbr})}, \rev{(\ref{defplus}), (\ref{defminus}),} (\ref{mu_tt_pm}), (\ref{mu_rt_pm}), (\ref{Lambda_tt}), and (\ref{Lambda_rt}), it applies for a suspension of our model microswimmers within a bulk viscous fluid in three spatial dimensions.

\section{Planar trapped microswimmer arrangements}
\label{numerics}


As a first application of the above DDFT, we are interested in the effect that the self-propulsion forces have on a confined assembly of microswimmers. In particular, this concerns the time evolution towards a final steady state when self-propulsion is suddenly switched on in an initially equilibrated system. Such a behavior could for instance be realized in experiments using light-activated microswimmers \cite{jiang2010active,volpe2011microswimmers,buttinoni2012active,buttinoni2013dynamical,palacci2013living, babel2014swimming,menzel2015focusing}. Here, we present numerical results for two-dimensional arrangements. That is, the density field $\rho^{(1)}(\mathbf r, \mathbf{\hat n},t)$ is calculated in the Cartesian $x$-$y$ plane, with the direction $\mathbf{\hat n}$ likewise confined to that plane and parameterized by one orientational angle. Concerning hydrodynamic interactions, the presence of a surrounding three-dimensional bulk fluid is still taken into account, as introduced in Sec.~\ref{Model}. Such a system could be realized approximately, for example, by confining the microswimmers to a plane using external laser potentials. Another realization could be microswimmers confined to the liquid--liquid interface between two immiscible fluids of identical viscosity.

The partial differential equation resulting from our DDFT, i.e.\ Eq.~(\ref{BBGKY1}) together with Eqs.~(\ref{DDFT_J1})--(\ref{DDFT_J6}), 
was discretized using a finite-difference scheme on a regular grid. The grid points were separated by distances $\Delta x=0.1$ in the spatial and $\Delta \phi=\pi/10$ in the angular direction, where we measure all lengths in units of $\sigma$. In each spatial direction, the numerical box length was $8$. 
$\rho^{(1)}(\mathbf r, \mathbf{\hat n},t)$ was iterated forward in time by employing a second-order Runge-Kutta scheme with fixed time step $\Delta t = 10^{-5}$. Here, we measure all times in units of the Brownian time scale 
$\tau_B=1/\mu^t$, where we recall that energies are \rev{given} in units of $k_BT$ (and lengths in units of $\sigma$). 
For simplicity and for practical purposes, periodic boundary conditions were used and the long-ranged hydrodynamic interactions were truncated at a cut-off radius of $r_c=1.875$.

Typically, self-propulsion is quantified by the P\'{e}clet number $\mathrm{Pe}$. Here, $\mathrm{Pe}$ corresponds to the ratio between the strength of self-propulsion and the strength of thermal fluctuations. In our units, we have $\mathrm{Pe}=|f|$. We choose fixed numerical values for all other system parameters, $a=L=0.75$, $\alpha=0.15$, $\epsilon_0=2$, and $k=30$, unless stated otherwise.

To study the time evolution of the confined system after switching on self-propulsion, we adhere to the following numerical protocol. 
First, we initialize the system by a random density profile and let it equilibrate with self-propulsion being switched off, i.e.\ $\mathrm{Pe}=f=0$. After equilibration, we turn on the active forces to $f\neq0$ and let the system find its new steady state, if existent, in non-equilibrium. 
Our results are presented in terms of the density profile 
\begin{equation}
\rho(\mathbf r,t) = \int \mathrm{d}\mathbf{\hat n}\, \rho^{(1)}(\mathbf r, \mathbf{\hat n}, t), 
\end{equation}
shown as color maps in the subsequent figures, as well as the orientational vector field 
\begin{equation}
\langle \mathbf{\hat n} \rangle (\mathbf r,t) = \int \mathrm{d}\mathbf{\hat n}\, \mathbf{\hat n}\,\rho^{(1)}(\mathbf r, \mathbf{\hat n}, t), 
\end{equation}
depicted as white arrows in the figures. 
In the following, we first describe our equilibrated initial state for $f=0$. Then we switch on self-propulsion to moderate values \rev{setting} $f\neq0$, but \rev{we} neglect hydrodynamic interactions between different swimmers. After that, we additionally include hydrodynamic interactions.

\begin{figure}
\includegraphics[width=8.5cm]{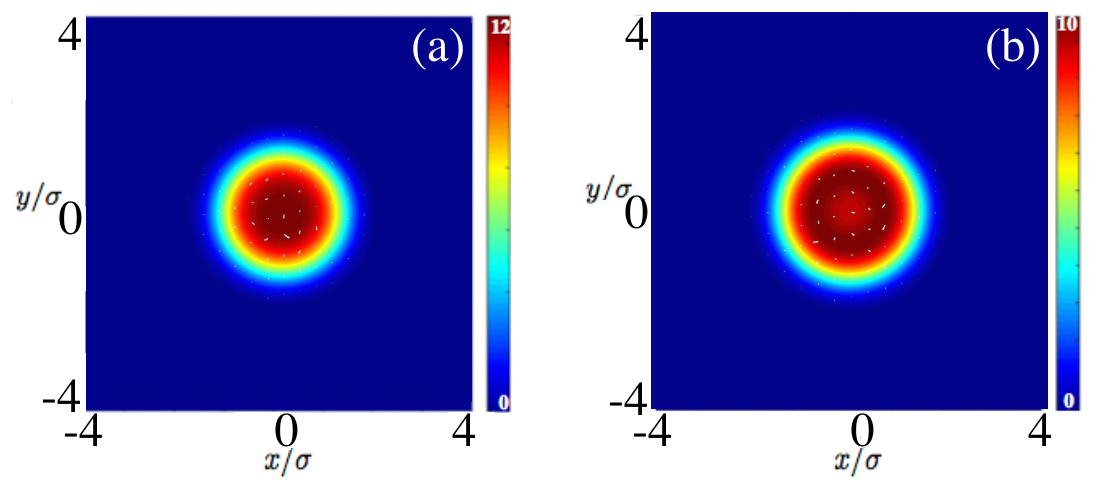}
\caption{Microswimmer density (color map) under confinement in equilibrium, i.e.\ for $\mathrm{Pe}=|f|=0$. In this situation, the density profile is rotationally symmetric, while the orientations are completely disordered. (a) Steric swimmer interactions switched off, $\epsilon_0=0$, showing a maximum density in the center of the confinement. (b) Steric swimmer interactions turned on, $\epsilon_0=2$, leading to a depletion of the swimmer density in the center.}
\label{Nfig1}
\end{figure}

First, for $f=0$, the system is in equilibrium. In our case, there are no orientation-dependent equilibrium interactions. Indeed, we find from Eqs.~(\ref{BBGKY1}) and (\ref{DDFT_J1})--(\ref{DDFT_J6}) that the swimmer orientations completely disorder. The equilibrium densities become independent of the swimmer orientations. Moreover, the system reaches a steady state, in which the entropic, steric inter-swimmer, and trapping forces balance each other. Hydrodynamic interactions do not affect these equilibrium states. As a result, the situation becomes rotationally symmetric in accordance with the rotational symmetry of the confinement. Fig.~\ref{Nfig1} shows two situations, one with the steric swimmer interactions switched off, $\epsilon_0=0$, see Fig.~\ref{Nfig1}(a), where the maximum swimmer density is found in the center of the confinement; and one with the steric interactions switched on, $\epsilon_0=2$, see Fig.~\ref{Nfig1}(b), which leads to a weak depletion of the density at the center point.

\begin{figure*}
\includegraphics[width=17.cm]{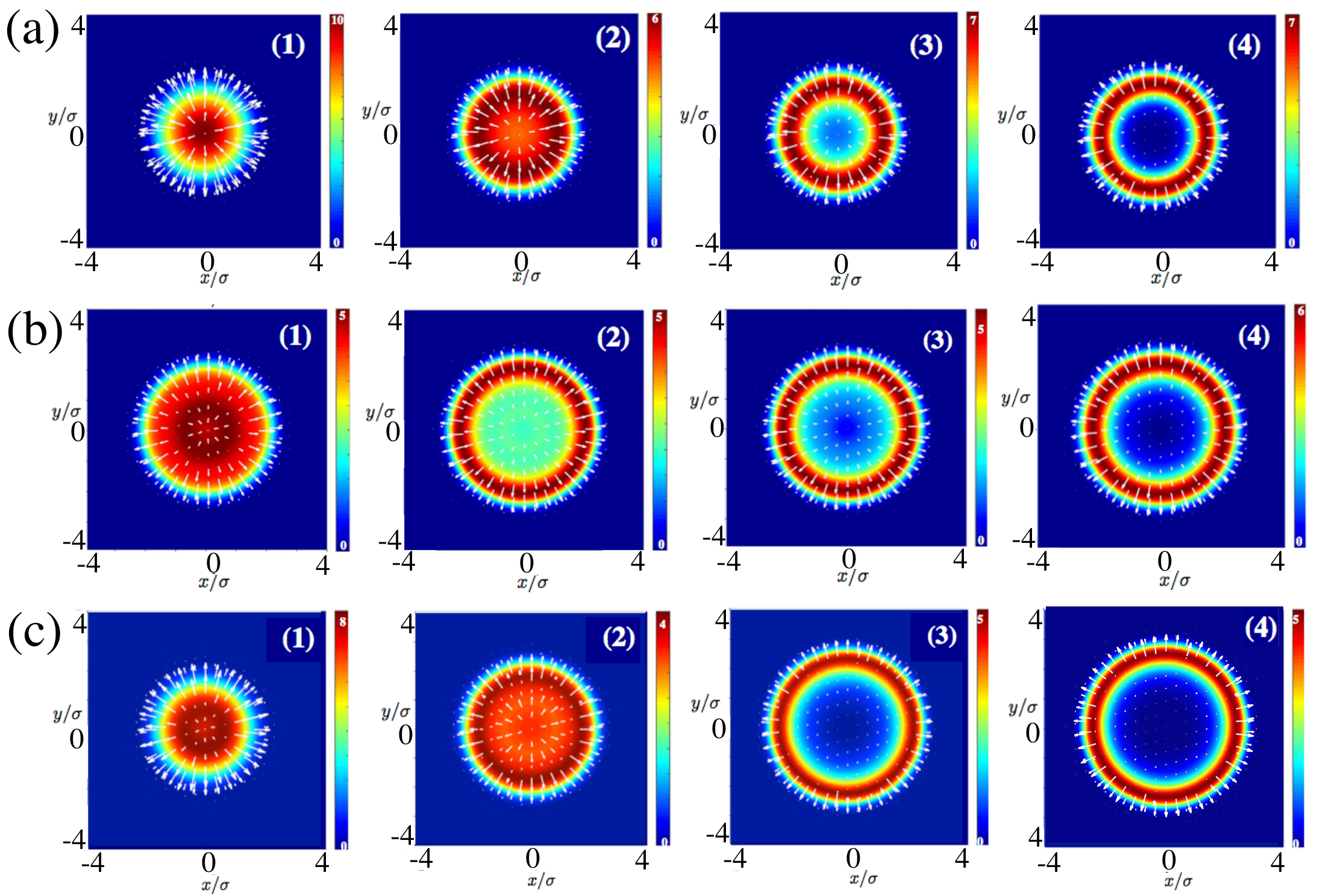}
\caption{
Time evolution of the density profiles (color maps) and orientation profiles (white arrows) of our confined microswimmer systems starting from the equilibrated states of $f=0$ depicted in Fig.~\ref{Nfig1}. At time $t=0$, the active force is switched on to $f=8$. (a) Snapshots without any steric ($\epsilon_0=0$) and without any hydrodynamic interactions between the swimmers at times $t=0.05$, $t=0.1$, $t=0.15$, and $t=0.4$. (b) Snapshots with steric ($\epsilon_0=2$) but still without any hydrodynamic interactions between the swimmers at times $t=0.02$, $t=0.06$, $t=0.08$, and $t=0.4$. (c) Snapshots with both steric ($\epsilon_0=2$) and hydrodynamic interactions between the swimmers at times $t=0.05$, $t=0.15$, $t=0.25$, and $t=0.4$.} 
\label{noint_steric_sterichydro}
\end{figure*}

We now turn on the active drive, $f\neq0$, yet to moderate magnitudes. Hydrodynamic interactions between different swimmers still remain switched off for the moment. Due to the active forces, the self-propelling microswimmers have an additional drive to work against the confining potential. In this way, they spread out and reach locations further separated from the center of the confinement. A time series is depicted in Fig.~\ref{noint_steric_sterichydro}(a),(b). 

Still, the situation apparently remains rotationally symmetric and finally reaches a steady state. Yet, the density in the center is now depleted, while a density ring forms at finite distance from the center as has been observed before in statistical and in particle-based approaches \cite{tailleur2009sedimentation,hennes2014self,menzel2015focusing}. From the white arrows in Fig.~\ref{noint_steric_sterichydro}, we find that the active forces drive the swimmers 
outwards against the confining potential barrier. In this sense, the potential blocks the swimmer motion in the final steady state \cite{enculescu2011active}. It takes a typical rotational diffusion time scale until a swimmer can reorient and leave the trapping location, before it propels towards another location on the high-density ring \cite{tailleur2009sedimentation,menzel2015focusing}.

The typical radius $\tilde{r}$ of the density ring in Fig.~\ref{noint_steric_sterichydro}(a), where different swimmers do not interact with each other, can readily be estimated. In this case, the $n$-swimmer densities for $n\geq2$ do not play a role. Consequently, in Eqs.~(\ref{eqJ1})--(\ref{eqJ6}) we find $\bm{\mathcal{ J}\!}_2=\bm{\mathcal{ J}\!}_4=\bm{\mathcal{ J}\!}_6=0$. The remaining orientational part in $\bm{\mathcal{ J}\!}_5$ decouples from the translational contributions and leads to free rotational diffusion. Finally, the remaining translational contributions in $\bm{\mathcal{ J}\!}_1$ and $\bm{\mathcal{ J}\!}_3$ must balance each other to allow for a steady state. This implies that the sum of the contributions from translational diffusion, confinement, and active forces must cancel. Assuming that at $r=\tilde{r}$ the density becomes maximum and exploiting the radial symmetry, we find 
\begin{equation}
\tilde{r}\approx\left|\frac{3g(\alpha)}{8}\right|^{1/3}\left|\frac{f}{k}\right|^{1/3},
\end{equation} 
where we have introduced the function 
\begin{equation}
g(\alpha)=\left(\frac{1-2\alpha}{\alpha(1-\alpha)}\right)\left(1-\frac{1-\alpha+\alpha^2}{3\alpha^2(1-\alpha)^2}\right)
\end{equation} 
for our special case of $a=L$. 
For harmonic confinement, this radius has been calculated in Refs.~\cite{hennes2014self, tailleur2009sedimentation}. It is conceivable that switching on an effective repulsion between the swimmers in the form of our soft steric interactions, $\epsilon_0=2$, adds to the spreading. This can be observed by the slightly larger diameter of the final density ring in Fig.~\ref{noint_steric_sterichydro}(b) when compared to the diameter in Fig.~\ref{noint_steric_sterichydro}(a).

In addition to the steric interactions between the microswimmers, we now also include the hydrodynamic interactions between them. At low to moderate magnitudes of the active forces, here $0 < \mathrm{Pe}=|f| \lesssim 10$, we still observe qualitatively the same scenario as described above in the absence of hydrodynamic interactions between the swimmers. At the end of our numerical simulation, see Fig.~\ref{noint_steric_sterichydro}(c), we again observe a density ring and a radial orientation of the swimmer axes. Due to the hydrodynamic interactions, however, the diameter of this density ring increases when compared to the case without hydrodynamic interactions between the swimmers, see the final states in Fig.~\ref{noint_steric_sterichydro}(b) and (c). Apparently, via the hydrodynamic interactions, the swimmers support each other in their collective propulsion against the confining potential. The presented snapshots were obtained for pushers ($f>0$), yet the results are qualitatively the same for pullers ($f<0$).

\begin{figure*}
\includegraphics[width=17.cm]{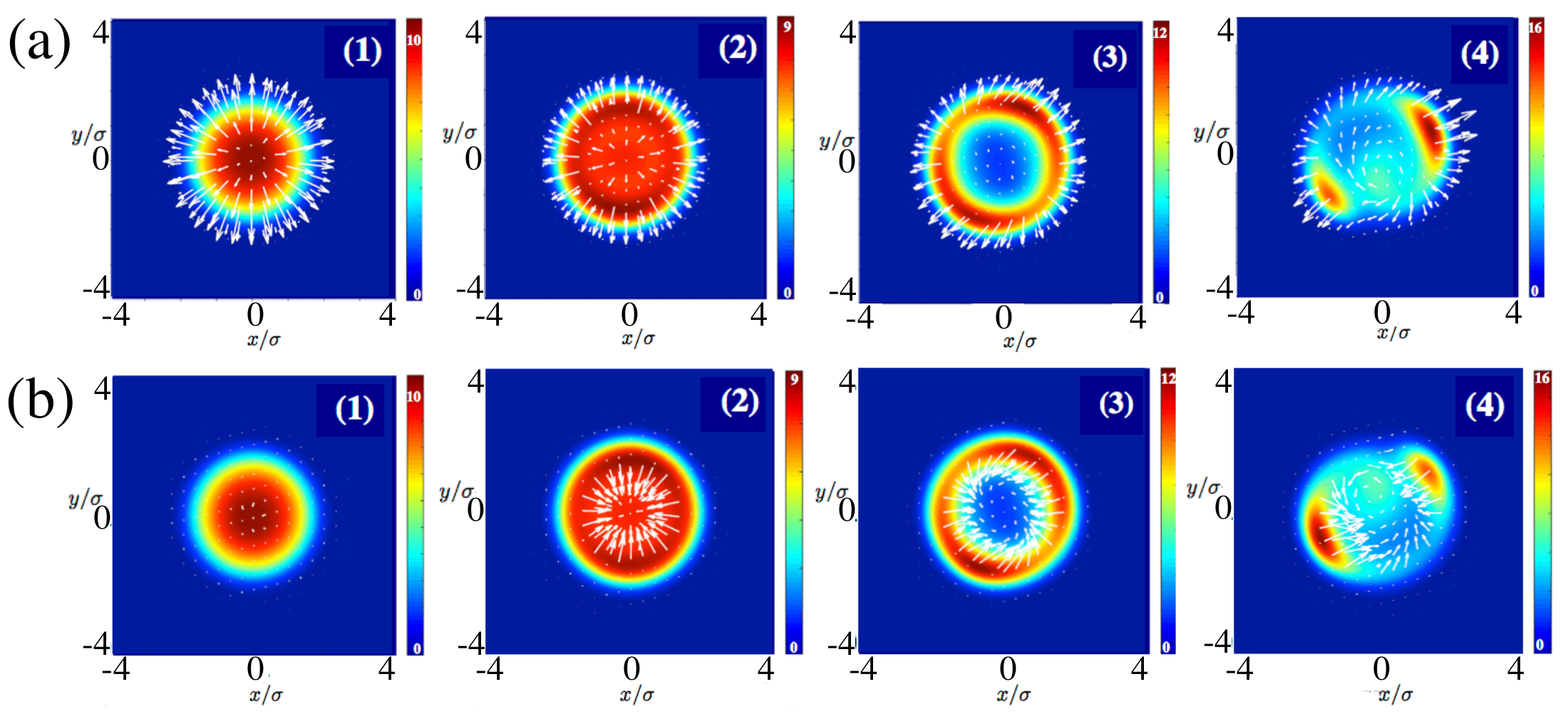}
\caption{
Time evolution of the density profiles (color maps) and orientation profiles (white arrows) of our microswimmer systems at (a) $f=50$ for pushers and (b) $f=-50$ for pullers. Both steric and hydrodynamic interactions between the swimmers are included. We observe rotational symmetry breaking within the plane. It corresponds to the formation of a ``hydrodynamic fluid pump'' consisting of self-assembled microswimmers. The snapshots were obtained at times $t=0.05$, $t=0.1$, $t=0.2$, and $t=0.8$.} 
\label{fig_pump}
\end{figure*}

From now on, we include both steric and hydrodynamic interactions between the microswimmers. We next consider increased values of the P\'eclet number of $10<\mathrm{Pe}=|f|\lesssim50$. When switching on this active force, the swimmers initially propel outwards from the center of the confinement as before. Although the system still appears to reach a steady state, the latter is not rotationally symmetric any more. We depict corresponding time evolutions in Fig.~\ref{fig_pump} for $f=\pm50$, i.e.\ for pushers and for pullers, respectively.

Pushers propel into the direction of the axis vector $\mathbf{\hat n}$, while pullers propel into the opposite direction, see Fig.~\ref{fig_microswimmer}. That is why the white arrows point outward in Fig.~\ref{fig_pump}(a) and inward in Fig.~\ref{fig_pump}(b). Since the rotational symmetry in the trapping plane is broken, a net fluid flow results in this plane. Therefore, the system can be viewed as a self-assembled ``hydrodynamic fluid pump'', which has been observed and interpreted before using particle-based lattice Boltzmann and Brownian dynamics simulations \cite{nash2010run, hennes2014self}.

\begin{figure}
\includegraphics[width=8.5cm]{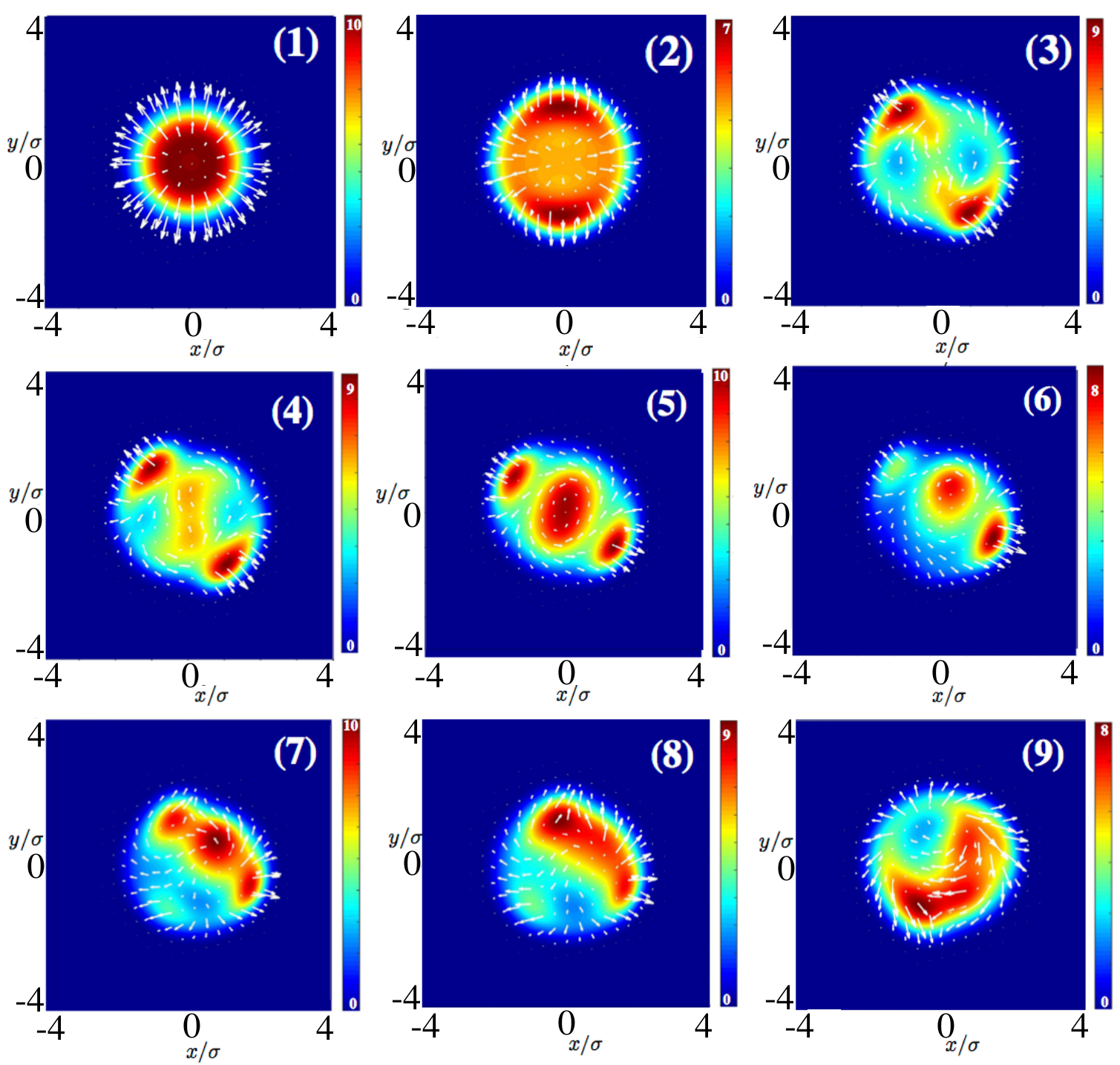}
\caption{Time evolution of the density profiles (color maps) and orientation profiles (white arrows) of pushers at $f=100$. Both steric and hydrodynamic interactions between the swimmers are included. This system does not reach a steady state any more within our numerically observed time window. The snapshots are obtained at times $t=0.02$, $t=0.1$, $t=0.25$, $t=0.3$, $t=1.25$, $t=2.5$, $t=2.7$, $t=3.0$, and $t=3.5$.} 
\label{unstable_pushers}
\end{figure}

\begin{figure}
\includegraphics[width=8.5cm]{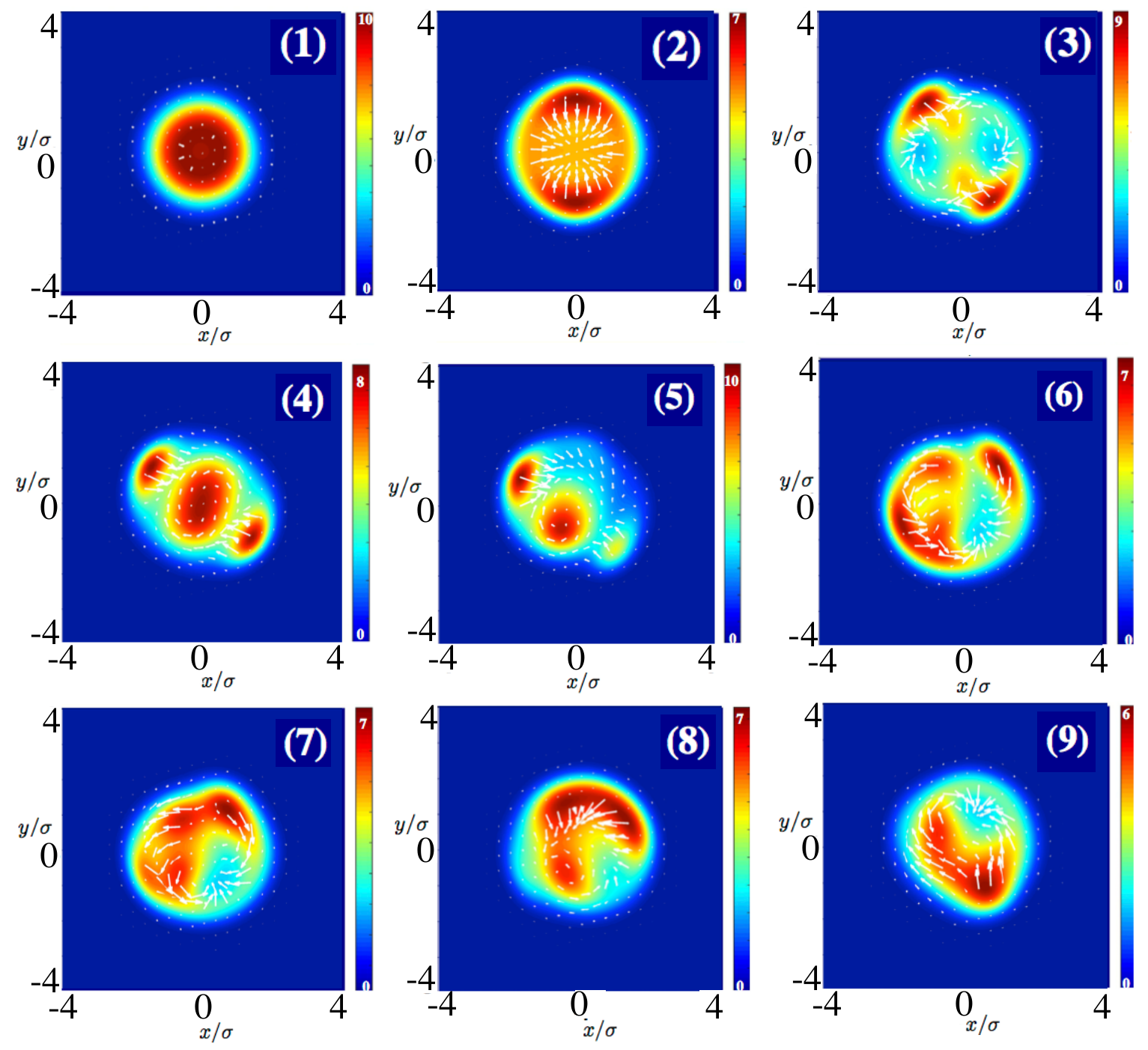}
\caption{Time evolution of the density profiles (color maps) and orientation profiles (white arrows) of pullers at $f=-100$. Both steric and hydrodynamic interactions between the swimmers are included. Again, this system does not reach a steady state any more within our numerically observed time window. The snapshots are obtained at times $t=0.02$, $t=0.1$, $t=0.25$, $t=0.3$, $t=1.25$, $t=2.5$, $t=2.7$, $t=3.0$, and $t=3.5$.} 
\label{unstable_pullers}
\end{figure}

Upon further increase of $\mathrm{Pe}=|f|$, the system does not enter a state of a steady hydrodynamic fluid pump any longer. Instead, the system becomes very dynamic. High density areas of localized orientational order of the swimmer axes form and continuously swap around within the spherical confinement. Examples for the time evolution are shown in Figs.~\ref{unstable_pushers} and \ref{unstable_pullers} for pushers and pullers, respectively. As far as we could test numerically, the system for these strong active forces does not reach a steady state any more. 

We briefly comment on the factors that lead to the observed destabilization effects. The first one breaks the initial rotational symmetry of Fig.~\ref{noint_steric_sterichydro}. It induces the formation of the hydrodynamic fluid pump, see Fig.~\ref{fig_pump}. 
In Refs.~\cite{nash2010run,hennes2014self} it was explained that rotational diffusion stabilizes the rotationally symmetric states of Fig.~\ref{noint_steric_sterichydro}. However, hydrodynamic interactions can lead to a destabilizing feedback mechanism that supports the rotational symmetry breaking. In brief, one has to realize that swimmers in the blocked state within the density ring transmit the confining forces to the surrounding fluid. As a consequence, fluid flows are induced. If a density fluctuation along the ring occurs, with a higher density at a certain spot, its induced fluid flow can reorient neighboring swimmers. The mechanism leads to positive feedback, i.e., the neighboring swimmers are reoriented such that they propel towards the high density region. In our formalism, a corresponding rotation--translation coupling to the influence of the confinement, introduced via $u_{ext}$, is contained in the current $\bm{\mathcal {J}\!}_4$ in Eq.~(\ref{DDFT_J4}).

The second destabilization occurs when at very high $\mathrm{Pe}=|f|$ a persistent hydrodynamic fluid pump as in Fig.~\ref{fig_pump} cannot be observed any more and the system becomes truly dynamic, see Figs.~\ref{unstable_pushers} and \ref{unstable_pullers}. This effect can be traced back to the rotation--translation coupling between swimmer rotations and the active point forces. Aligned 
and concentrated active forces can induce rotations of neighboring swimmer bodies, which in turn can lead to rotational instabilities. This effect is proportional to the strength of the active forces $|f|$. At high $\mathrm{Pe}=|f|$, it apparently cannot be stabilized any longer. In our formalism, this contribution is represented by the current $\bm{\mathcal {J}\!}_6$ in Eq.~(\ref{DDFT_J6}). We have numerically tested our assertion by deactivating this current.

\section{Conclusions}
\label{conclusion}

In this work, we have derived a statistical characterization of dilute to semi-dilute suspensions of identical self-propelled microswimmers in the form of a dynamical density functional theory (DDFT). Our simple model microswimmers consist of a body that experiences hydrodynamic drag from the surrounding fluid, plus two separated active point-like force centers. Two antiparallel active point forces of equal magnitude are exerted by these force centers onto the surrounding fluid and set it into motion. Pushing and pulling swimming mechanisms can easily be distinguished. We include both hydrodynamic and steric interactions between the swimmers, as well as the effect of an external trapping potential. Hydrodynamic interactions result both from the active forces as well as from steric and external forces acting on the swimmer bodies. 
At this time, axially symmetric model microswimmers are considered, thus active torques do not arise. Moreover, only isotropic steric interactions are taken into account.

Our DDFT describes the overdamped time evolution of the microswimmer density, both concerning positions and orientations of the swimmers. As a first application and test of the theory, we consider a crowd of microswimmers restricted to planar motion within a three-dimensional bulk fluid. Such an arrangement could be achieved, for instance, using external trapping laser potentials, or by confining the swimmers to an interface between two immiscible fluids of equal viscosity. Moreover, an additional radially symmetric trapping potential was taken into account. Within this framework, the theory was evaluated numerically.

The numerical calculations started from an initial state in which a crowd of microswimmers is concentrated in the center of the spherical trap. At low P\'eclet numbers, which means low magnitude of the active forces, the microswimmers propel outwards, where in a final stationary state they form a ring-like density profile. This effect remains when hydrodynamic interactions are switched off in the numerical calculations \rev{as reported in different frameworks previously \cite{tailleur2009sedimentation, pototsky2012active, menzel2015focusing}}. Increasing the P\'eclet number and including hydrodynamic interactions, the numerical evaluation of the DDFT shows a breaking of rotational symmetry. The ring-like density profile observed for lower P\'eclet numbers now is replaced by concentrated density spots. Likewise, this effect has been observed before by different approaches, both for lattice Boltzmann as well as for Brownian dynamics simulations \cite{nash2010run, hennes2014self}. Due to the polar order of the swimmers within the concentrated spots and the resulting fluid flows, this state was identified as a hydrodynamic fluid pump. Obviously, our DDFT reproduces these previously identified effects, \rev{which stresses its potential}. Finally, upon further increase of the P\'eclet number, the numerical evaluation shows a persistently dynamic state of migrating density clouds. 

\rev{As common for DDFT approaches, our description partially leans on equilibrium concepts. However, the situation under consideration is an intrinsically non-equilibrium one. 
For instance, we used a temperature variable to measure energies and to define the P\'eclet number. We identified this variable with a constant temperature of the background fluid. It might be stabilized by coupling to an external heat bath. Yet, strictly speaking, the energy input due to self-propulsion can lead to local changes in the temperature. This issue may become relevant for thermally driven artificial microswimmers in the form of externally heated Janus particles \cite{jiang2010active,volpe2011microswimmers,buttinoni2012active,buttinoni2013dynamical}. Moreover, in different contexts, effective temperatures were introduced to correctly describe deviations from equilibrium temperatures in driven systems \cite{ono2002effective,seifert2012stochastic}. In the present case, motion is induced by the microswimmers in the surrounding fluid.  This issue may be investigated in a profound analysis, but is not addressed here. 
We only remark that the translational and rotational diffusion behavior [represented by the terms containing $\ln P$ in Eqs.~(\ref{force}) and (\ref{torque})] may need to be modified if local deviations from the heat bath temperature become perceptible. Our framework of hydrodynamic interactions remains basically unaffected, as long as local deviations of the viscosity or density remain negligible.}

\rev{In the derivation of the statistical theory, conservation of the probability to find the particles somewhere in phase space [Eq.~(\ref{Smoluchowski})] remains, of course, unaltered by the non-equilibrium nature of our system. Therefore, apart from the points mentioned above, no equilibrium approximations are involved in our initial statistical equations [Eqs.~(\ref{Smoluchowski})--(\ref{eqJ6})]. The situation changes when formulae that were derived exactly in the context of equilibrium DFT are adapted [Eqs.~(\ref{Omega})--(\ref{DDFT_J6})] to close our hierarchy of non-equilibrium statistical equations. This crucial step is generic for DDFTs but needs to be tested by numerical evaluation of the full statistical equations or by particle-based simulations. In our case, we do reproduce corresponding results of previous particle-based simulations. This stresses the power of our newly derived DDFT in describing the complex behavior of microswimmer suspensions. As a side remark, we note that mainly the steric inter-particle interactions are directly concerned by the DFT approximation [see the presence of the $u(\mathbf{r},\mathbf{r}')$ terms on the left-hand sides of Eqs.~(\ref{YBG1}) and (\ref{YBG2})]. Further analysis may be necessary when such interactions form the central focus of a quantitative DDFT approach.} 

Naturally, future applications and extensions of our theory are manifold. It should be further compared to particle-based simulations and possible experiments to learn more about the range of its predictive power. As indicated above, an obvious next step is to extend the theory to include active torques and anisotropic steric interactions. Moreover, the influence of different effective steric interactions, for instance hard-body interactions, may be investigated \cite{wensink2013differently}. 
Other variations include, for example, the hydrodynamic effect of confining boundaries \cite{spagnolie2012hydrodynamics} or external magnetic alignment fields acting onto magnetic microswimmers \cite{baraban2012catalytic}. 
In the longer term, an extension of the investigations to denser crystalized systems as well as three-dimensional numerical implementations are desirable.

\begin{acknowledgments}

The authors thank the Deutsche Forschungsgemeinschaft for support of this work through the SPP 1726 on ``Microswimmers''. 

\end{acknowledgments}

\subsection*{Author Contributions}
A.M.M., C.H., and H.L.\ developed the theory. A.S.\ performed the numerical evaluations. A.M.M., A.S., and H.L.\ wrote the paper. 


\begin{thebibliography}{117}
\expandafter\ifx\csname natexlab\endcsname\relax\def\natexlab#1{#1}\fi
\expandafter\ifx\csname bibnamefont\endcsname\relax
  \def\bibnamefont#1{#1}\fi
\expandafter\ifx\csname bibfnamefont\endcsname\relax
  \def\bibfnamefont#1{#1}\fi
\expandafter\ifx\csname citenamefont\endcsname\relax
  \def\citenamefont#1{#1}\fi
\expandafter\ifx\csname url\endcsname\relax
  \def\url#1{\texttt{#1}}\fi
\expandafter\ifx\csname urlprefix\endcsname\relax\def\urlprefix{URL }\fi
\providecommand{\bibinfo}[2]{#2}
\providecommand{\eprint}[2][]{\url{#2}}

\bibitem[{\citenamefont{Lauga and Powers}(2009)}]{lauga2009hydrodynamics}
\bibinfo{author}{\bibfnamefont{E.}~\bibnamefont{Lauga}} \bibnamefont{and}
  \bibinfo{author}{\bibfnamefont{T.~R.} \bibnamefont{Powers}},
  \bibinfo{journal}{Rep. Prog. Phys.} \textbf{\bibinfo{volume}{72}},
  \bibinfo{pages}{096601} (\bibinfo{year}{2009}).

\bibitem[{\citenamefont{Spagnolie and
  Lauga}(2012)}]{spagnolie2012hydrodynamics}
\bibinfo{author}{\bibfnamefont{S.~E.} \bibnamefont{Spagnolie}}
  \bibnamefont{and} \bibinfo{author}{\bibfnamefont{E.}~\bibnamefont{Lauga}},
  \bibinfo{journal}{J. Fluid. Mech.} \textbf{\bibinfo{volume}{700}},
  \bibinfo{pages}{105} (\bibinfo{year}{2012}).

\bibitem[{\citenamefont{Menzel}(2015{\natexlab{a}})}]{menzel2015tuned}
\bibinfo{author}{\bibfnamefont{A.~M.} \bibnamefont{Menzel}},
  \bibinfo{journal}{Phys. Rep.} \textbf{\bibinfo{volume}{554}},
  \bibinfo{pages}{1} (\bibinfo{year}{2015}{\natexlab{a}}).

\bibitem[{\citenamefont{Elgeti et~al.}(2015)\citenamefont{Elgeti, Winkler, and
  Gompper}}]{elgeti2015physics}
\bibinfo{author}{\bibfnamefont{J.}~\bibnamefont{Elgeti}},
  \bibinfo{author}{\bibfnamefont{R.~G.} \bibnamefont{Winkler}},
  \bibnamefont{and} \bibinfo{author}{\bibfnamefont{G.}~\bibnamefont{Gompper}},
  \bibinfo{journal}{Rep. Prog. Phys.} \textbf{\bibinfo{volume}{78}},
  \bibinfo{pages}{056601} (\bibinfo{year}{2015}).

\bibitem[{\citenamefont{Eisenbach and Giojalas}(2006)}]{eisenbach2006sperm}
\bibinfo{author}{\bibfnamefont{M.}~\bibnamefont{Eisenbach}} \bibnamefont{and}
  \bibinfo{author}{\bibfnamefont{L.~C.} \bibnamefont{Giojalas}},
  \bibinfo{journal}{Nature Rev. Mol. Cell Biol.} \textbf{\bibinfo{volume}{7}},
  \bibinfo{pages}{276} (\bibinfo{year}{2006}).

\bibitem[{\citenamefont{Berg and Brown}(1972)}]{berg1972chemotaxis}
\bibinfo{author}{\bibfnamefont{H.~C.} \bibnamefont{Berg}} \bibnamefont{and}
  \bibinfo{author}{\bibfnamefont{D.~A.} \bibnamefont{Brown}},
  \bibinfo{journal}{Nature} \textbf{\bibinfo{volume}{239}},
  \bibinfo{pages}{500} (\bibinfo{year}{1972}).

\bibitem[{\citenamefont{Walther and M\"uller}(2013)}]{walther2013janus}
\bibinfo{author}{\bibfnamefont{A.}~\bibnamefont{Walther}} \bibnamefont{and}
  \bibinfo{author}{\bibfnamefont{A.~H.~E.} \bibnamefont{M\"uller}},
  \bibinfo{journal}{Chem. Rev.} \textbf{\bibinfo{volume}{113}},
  \bibinfo{pages}{5194} (\bibinfo{year}{2013}).

\bibitem[{\citenamefont{Romanczuk et~al.}(2012)\citenamefont{Romanczuk,
  B{\"a}r, Ebeling, Lindner, and Schimansky-Geier}}]{romanczuk2012active}
\bibinfo{author}{\bibfnamefont{P.}~\bibnamefont{Romanczuk}},
  \bibinfo{author}{\bibfnamefont{M.}~\bibnamefont{B{\"a}r}},
  \bibinfo{author}{\bibfnamefont{W.}~\bibnamefont{Ebeling}},
  \bibinfo{author}{\bibfnamefont{B.}~\bibnamefont{Lindner}}, \bibnamefont{and}
  \bibinfo{author}{\bibfnamefont{L.}~\bibnamefont{Schimansky-Geier}},
  \bibinfo{journal}{Eur. Phys. J. Special Topics}
  \textbf{\bibinfo{volume}{202}}, \bibinfo{pages}{1} (\bibinfo{year}{2012}).

\bibitem[{\citenamefont{Cates}(2012)}]{cates2012diffusive}
\bibinfo{author}{\bibfnamefont{M.~E.} \bibnamefont{Cates}},
  \bibinfo{journal}{Rep. Prog. Phys.} \textbf{\bibinfo{volume}{75}},
  \bibinfo{pages}{042601} (\bibinfo{year}{2012}).

\bibitem[{\citenamefont{Marchetti et~al.}(2013)\citenamefont{Marchetti, Joanny,
  Ramaswamy, Liverpool, Prost, Rao, and Simha}}]{marchetti2013hydrodynamics}
\bibinfo{author}{\bibfnamefont{M.~C.} \bibnamefont{Marchetti}},
  \bibinfo{author}{\bibfnamefont{J.~F.} \bibnamefont{Joanny}},
  \bibinfo{author}{\bibfnamefont{S.}~\bibnamefont{Ramaswamy}},
  \bibinfo{author}{\bibfnamefont{T.~B.} \bibnamefont{Liverpool}},
  \bibinfo{author}{\bibfnamefont{J.}~\bibnamefont{Prost}},
  \bibinfo{author}{\bibfnamefont{M.}~\bibnamefont{Rao}}, \bibnamefont{and}
  \bibinfo{author}{\bibfnamefont{R.~A.} \bibnamefont{Simha}},
  \bibinfo{journal}{Rev. Mod. Phys.} \textbf{\bibinfo{volume}{85}},
  \bibinfo{pages}{1143} (\bibinfo{year}{2013}).

\bibitem[{\citenamefont{Vicsek et~al.}(1995)\citenamefont{Vicsek, Czir{\'o}k,
  Ben-Jacob, Cohen, and Shochet}}]{vicsek1995novel}
\bibinfo{author}{\bibfnamefont{T.}~\bibnamefont{Vicsek}},
  \bibinfo{author}{\bibfnamefont{A.}~\bibnamefont{Czir{\'o}k}},
  \bibinfo{author}{\bibfnamefont{E.}~\bibnamefont{Ben-Jacob}},
  \bibinfo{author}{\bibfnamefont{I.}~\bibnamefont{Cohen}}, \bibnamefont{and}
  \bibinfo{author}{\bibfnamefont{O.}~\bibnamefont{Shochet}},
  \bibinfo{journal}{Phys. Rev. Lett.} \textbf{\bibinfo{volume}{75}},
  \bibinfo{pages}{1226} (\bibinfo{year}{1995}).

\bibitem[{\citenamefont{Toner and Tu}(1995)}]{toner1995long}
\bibinfo{author}{\bibfnamefont{J.}~\bibnamefont{Toner}} \bibnamefont{and}
  \bibinfo{author}{\bibfnamefont{Y.}~\bibnamefont{Tu}}, \bibinfo{journal}{Phys.
  Rev. Lett.} \textbf{\bibinfo{volume}{75}}, \bibinfo{pages}{4326}
  (\bibinfo{year}{1995}).

\bibitem[{\citenamefont{Toner and Tu}(1998)}]{toner1998flocks}
\bibinfo{author}{\bibfnamefont{J.}~\bibnamefont{Toner}} \bibnamefont{and}
  \bibinfo{author}{\bibfnamefont{Y.}~\bibnamefont{Tu}}, \bibinfo{journal}{Phys.
  Rev. E} \textbf{\bibinfo{volume}{58}}, \bibinfo{pages}{4828}
  (\bibinfo{year}{1998}).

\bibitem[{\citenamefont{Toner et~al.}(2005)\citenamefont{Toner, Tu, and
  Ramaswamy}}]{toner2005hydrodynamics}
\bibinfo{author}{\bibfnamefont{J.}~\bibnamefont{Toner}},
  \bibinfo{author}{\bibfnamefont{Y.}~\bibnamefont{Tu}}, \bibnamefont{and}
  \bibinfo{author}{\bibfnamefont{S.}~\bibnamefont{Ramaswamy}},
  \bibinfo{journal}{Ann. Phys.} \textbf{\bibinfo{volume}{318}},
  \bibinfo{pages}{170} (\bibinfo{year}{2005}).

\bibitem[{\citenamefont{Bertin et~al.}(2006)\citenamefont{Bertin, Droz, and
  Gr{\'e}goire}}]{bertin2006boltzmann}
\bibinfo{author}{\bibfnamefont{E.}~\bibnamefont{Bertin}},
  \bibinfo{author}{\bibfnamefont{M.}~\bibnamefont{Droz}}, \bibnamefont{and}
  \bibinfo{author}{\bibfnamefont{G.}~\bibnamefont{Gr{\'e}goire}},
  \bibinfo{journal}{Phys. Rev. E} \textbf{\bibinfo{volume}{74}},
  \bibinfo{pages}{022101} (\bibinfo{year}{2006}).

\bibitem[{\citenamefont{Bertin et~al.}(2009)\citenamefont{Bertin, Droz, and
  Gr{\'e}goire}}]{bertin2009hydrodynamic}
\bibinfo{author}{\bibfnamefont{E.}~\bibnamefont{Bertin}},
  \bibinfo{author}{\bibfnamefont{M.}~\bibnamefont{Droz}}, \bibnamefont{and}
  \bibinfo{author}{\bibfnamefont{G.}~\bibnamefont{Gr{\'e}goire}},
  \bibinfo{journal}{J. Phys. A: Math. Theor.} \textbf{\bibinfo{volume}{42}},
  \bibinfo{pages}{445001} (\bibinfo{year}{2009}).

\bibitem[{\citenamefont{Leoni and Liverpool}(2010)}]{leoni2010swimmers}
\bibinfo{author}{\bibfnamefont{M.}~\bibnamefont{Leoni}} \bibnamefont{and}
  \bibinfo{author}{\bibfnamefont{T.~B.} \bibnamefont{Liverpool}},
  \bibinfo{journal}{Phys. Rev. Lett.} \textbf{\bibinfo{volume}{105}},
  \bibinfo{pages}{238102} (\bibinfo{year}{2010}).

\bibitem[{\citenamefont{Gr{\'e}goire and Chat{\'e}}(2004)}]{gregoire2004onset}
\bibinfo{author}{\bibfnamefont{G.}~\bibnamefont{Gr{\'e}goire}}
  \bibnamefont{and}
  \bibinfo{author}{\bibfnamefont{H.}~\bibnamefont{Chat{\'e}}},
  \bibinfo{journal}{Phys. Rev. Lett.} \textbf{\bibinfo{volume}{92}},
  \bibinfo{pages}{025702} (\bibinfo{year}{2004}).

\bibitem[{\citenamefont{Mishra et~al.}(2010)\citenamefont{Mishra, Baskaran, and
  Marchetti}}]{mishra2010fluctuations}
\bibinfo{author}{\bibfnamefont{S.}~\bibnamefont{Mishra}},
  \bibinfo{author}{\bibfnamefont{A.}~\bibnamefont{Baskaran}}, \bibnamefont{and}
  \bibinfo{author}{\bibfnamefont{M.~C.} \bibnamefont{Marchetti}},
  \bibinfo{journal}{Phys. Rev. E} \textbf{\bibinfo{volume}{81}},
  \bibinfo{pages}{061916} (\bibinfo{year}{2010}).

\bibitem[{\citenamefont{Schaller et~al.}(2010)\citenamefont{Schaller, Weber,
  Semmrich, Frey, and Bausch}}]{schaller2010polar}
\bibinfo{author}{\bibfnamefont{V.}~\bibnamefont{Schaller}},
  \bibinfo{author}{\bibfnamefont{C.}~\bibnamefont{Weber}},
  \bibinfo{author}{\bibfnamefont{C.}~\bibnamefont{Semmrich}},
  \bibinfo{author}{\bibfnamefont{E.}~\bibnamefont{Frey}}, \bibnamefont{and}
  \bibinfo{author}{\bibfnamefont{A.~R.} \bibnamefont{Bausch}},
  \bibinfo{journal}{Nature} \textbf{\bibinfo{volume}{467}}, \bibinfo{pages}{73}
  (\bibinfo{year}{2010}).

\bibitem[{\citenamefont{Menzel}(2012)}]{menzel2012collective}
\bibinfo{author}{\bibfnamefont{A.~M.} \bibnamefont{Menzel}},
  \bibinfo{journal}{Phys. Rev. E} \textbf{\bibinfo{volume}{85}},
  \bibinfo{pages}{021912} (\bibinfo{year}{2012}).

\bibitem[{\citenamefont{Ihle}(2013)}]{ihle2013invasion}
\bibinfo{author}{\bibfnamefont{T.}~\bibnamefont{Ihle}}, \bibinfo{journal}{Phys.
  Rev. E} \textbf{\bibinfo{volume}{88}}, \bibinfo{pages}{040303}
  (\bibinfo{year}{2013}).

\bibitem[{\citenamefont{Caussin et~al.}(2014)\citenamefont{Caussin, Solon,
  Peshkov, Chat{\'e}, Dauxois, Tailleur, Vitelli, and
  Bartolo}}]{caussin2014emergent}
\bibinfo{author}{\bibfnamefont{J.-B.} \bibnamefont{Caussin}},
  \bibinfo{author}{\bibfnamefont{A.}~\bibnamefont{Solon}},
  \bibinfo{author}{\bibfnamefont{A.}~\bibnamefont{Peshkov}},
  \bibinfo{author}{\bibfnamefont{H.}~\bibnamefont{Chat{\'e}}},
  \bibinfo{author}{\bibfnamefont{T.}~\bibnamefont{Dauxois}},
  \bibinfo{author}{\bibfnamefont{J.}~\bibnamefont{Tailleur}},
  \bibinfo{author}{\bibfnamefont{V.}~\bibnamefont{Vitelli}}, \bibnamefont{and}
  \bibinfo{author}{\bibfnamefont{D.}~\bibnamefont{Bartolo}},
  \bibinfo{journal}{Phys. Rev. Lett.} \textbf{\bibinfo{volume}{112}},
  \bibinfo{pages}{148102} (\bibinfo{year}{2014}).

\bibitem[{\citenamefont{Ohta and Yamanaka}(2014)}]{ohta2014soliton}
\bibinfo{author}{\bibfnamefont{T.}~\bibnamefont{Ohta}} \bibnamefont{and}
  \bibinfo{author}{\bibfnamefont{S.}~\bibnamefont{Yamanaka}},
  \bibinfo{journal}{Prog. Theor. Exp. Phys.} \textbf{\bibinfo{volume}{2014}},
  \bibinfo{pages}{011J01} (\bibinfo{year}{2014}).

\bibitem[{\citenamefont{Wensink et~al.}(2012)\citenamefont{Wensink, Dunkel,
  Heidenreich, Drescher, Goldstein, L{\"o}wen, and Yeomans}}]{wensink2012meso}
\bibinfo{author}{\bibfnamefont{H.~H.} \bibnamefont{Wensink}},
  \bibinfo{author}{\bibfnamefont{J.}~\bibnamefont{Dunkel}},
  \bibinfo{author}{\bibfnamefont{S.}~\bibnamefont{Heidenreich}},
  \bibinfo{author}{\bibfnamefont{K.}~\bibnamefont{Drescher}},
  \bibinfo{author}{\bibfnamefont{R.~E.} \bibnamefont{Goldstein}},
  \bibinfo{author}{\bibfnamefont{H.}~\bibnamefont{L{\"o}wen}},
  \bibnamefont{and} \bibinfo{author}{\bibfnamefont{J.~M.}
  \bibnamefont{Yeomans}}, \bibinfo{journal}{Proc. Natl. Acad. Sci. USA}
  \textbf{\bibinfo{volume}{109}}, \bibinfo{pages}{14308}
  (\bibinfo{year}{2012}).

\bibitem[{\citenamefont{S{\l}omka and Dunkel}(2015)}]{slomka2015generalized}
\bibinfo{author}{\bibfnamefont{J.}~\bibnamefont{S{\l}omka}} \bibnamefont{and}
  \bibinfo{author}{\bibfnamefont{J.}~\bibnamefont{Dunkel}},
  \bibinfo{journal}{Eur. Phys. J. Special Topics}
  \textbf{\bibinfo{volume}{224}}, \bibinfo{pages}{1349} (\bibinfo{year}{2015}).

\bibitem[{\citenamefont{Ordemann et~al.}(2003)\citenamefont{Ordemann, Balazsi,
  and Moss}}]{ordemann2003pattern}
\bibinfo{author}{\bibfnamefont{A.}~\bibnamefont{Ordemann}},
  \bibinfo{author}{\bibfnamefont{G.}~\bibnamefont{Balazsi}}, \bibnamefont{and}
  \bibinfo{author}{\bibfnamefont{F.}~\bibnamefont{Moss}},
  \bibinfo{journal}{Physica A} \textbf{\bibinfo{volume}{325}},
  \bibinfo{pages}{260} (\bibinfo{year}{2003}).

\bibitem[{\citenamefont{Peruani et~al.}(2006)\citenamefont{Peruani, Deutsch,
  and B{\"a}r}}]{peruani2006nonequilibrium}
\bibinfo{author}{\bibfnamefont{F.}~\bibnamefont{Peruani}},
  \bibinfo{author}{\bibfnamefont{A.}~\bibnamefont{Deutsch}}, \bibnamefont{and}
  \bibinfo{author}{\bibfnamefont{M.}~\bibnamefont{B{\"a}r}},
  \bibinfo{journal}{Phys. Rev. E} \textbf{\bibinfo{volume}{74}},
  \bibinfo{pages}{030904} (\bibinfo{year}{2006}).

\bibitem[{\citenamefont{Ishikawa and Pedley}(2008)}]{ishikawa2008coherent}
\bibinfo{author}{\bibfnamefont{T.}~\bibnamefont{Ishikawa}} \bibnamefont{and}
  \bibinfo{author}{\bibfnamefont{T.~J.} \bibnamefont{Pedley}},
  \bibinfo{journal}{Phys. Rev. Lett.} \textbf{\bibinfo{volume}{100}},
  \bibinfo{pages}{088103} (\bibinfo{year}{2008}).

\bibitem[{\citenamefont{Tailleur and Cates}(2008)}]{tailleur2008statistical}
\bibinfo{author}{\bibfnamefont{J.}~\bibnamefont{Tailleur}} \bibnamefont{and}
  \bibinfo{author}{\bibfnamefont{M.~E.} \bibnamefont{Cates}},
  \bibinfo{journal}{Phys. Rev. Lett.} \textbf{\bibinfo{volume}{100}},
  \bibinfo{pages}{218103} (\bibinfo{year}{2008}).

\bibitem[{\citenamefont{Theurkauff et~al.}(2012)\citenamefont{Theurkauff,
  Cottin-Bizonne, Palacci, Ybert, and Bocquet}}]{theurkauff2012dynamic}
\bibinfo{author}{\bibfnamefont{I.}~\bibnamefont{Theurkauff}},
  \bibinfo{author}{\bibfnamefont{C.}~\bibnamefont{Cottin-Bizonne}},
  \bibinfo{author}{\bibfnamefont{J.}~\bibnamefont{Palacci}},
  \bibinfo{author}{\bibfnamefont{C.}~\bibnamefont{Ybert}}, \bibnamefont{and}
  \bibinfo{author}{\bibfnamefont{L.}~\bibnamefont{Bocquet}},
  \bibinfo{journal}{Phys. Rev. Lett.} \textbf{\bibinfo{volume}{108}},
  \bibinfo{pages}{268303} (\bibinfo{year}{2012}).

\bibitem[{\citenamefont{Fily and Marchetti}(2012)}]{fily2012athermal}
\bibinfo{author}{\bibfnamefont{Y.}~\bibnamefont{Fily}} \bibnamefont{and}
  \bibinfo{author}{\bibfnamefont{M.~C.} \bibnamefont{Marchetti}},
  \bibinfo{journal}{Phys. Rev. Lett.} \textbf{\bibinfo{volume}{108}},
  \bibinfo{pages}{235702} (\bibinfo{year}{2012}).

\bibitem[{\citenamefont{Cates and Tailleur}(2013)}]{cates2013active}
\bibinfo{author}{\bibfnamefont{M.~E.} \bibnamefont{Cates}} \bibnamefont{and}
  \bibinfo{author}{\bibfnamefont{J.}~\bibnamefont{Tailleur}},
  \bibinfo{journal}{EPL (Europhys. Lett.)} \textbf{\bibinfo{volume}{101}},
  \bibinfo{pages}{20010} (\bibinfo{year}{2013}).

\bibitem[{\citenamefont{Redner et~al.}(2013)\citenamefont{Redner, Hagan, and
  Baskaran}}]{redner2013structure}
\bibinfo{author}{\bibfnamefont{G.~S.} \bibnamefont{Redner}},
  \bibinfo{author}{\bibfnamefont{M.~F.} \bibnamefont{Hagan}}, \bibnamefont{and}
  \bibinfo{author}{\bibfnamefont{A.}~\bibnamefont{Baskaran}},
  \bibinfo{journal}{Phys. Rev. Lett.} \textbf{\bibinfo{volume}{110}},
  \bibinfo{pages}{055701} (\bibinfo{year}{2013}).

\bibitem[{\citenamefont{Palacci et~al.}(2013)\citenamefont{Palacci, Sacanna,
  Steinberg, Pine, and Chaikin}}]{palacci2013living}
\bibinfo{author}{\bibfnamefont{J.}~\bibnamefont{Palacci}},
  \bibinfo{author}{\bibfnamefont{S.}~\bibnamefont{Sacanna}},
  \bibinfo{author}{\bibfnamefont{A.~P.} \bibnamefont{Steinberg}},
  \bibinfo{author}{\bibfnamefont{D.~J.} \bibnamefont{Pine}}, \bibnamefont{and}
  \bibinfo{author}{\bibfnamefont{P.~M.} \bibnamefont{Chaikin}},
  \bibinfo{journal}{Science} \textbf{\bibinfo{volume}{339}},
  \bibinfo{pages}{936} (\bibinfo{year}{2013}).

\bibitem[{\citenamefont{Buttinoni et~al.}(2013)\citenamefont{Buttinoni,
  Bialk{\'e}, K{\"u}mmel, L{\"o}wen, Bechinger, and
  Speck}}]{buttinoni2013dynamical}
\bibinfo{author}{\bibfnamefont{I.}~\bibnamefont{Buttinoni}},
  \bibinfo{author}{\bibfnamefont{J.}~\bibnamefont{Bialk{\'e}}},
  \bibinfo{author}{\bibfnamefont{F.}~\bibnamefont{K{\"u}mmel}},
  \bibinfo{author}{\bibfnamefont{H.}~\bibnamefont{L{\"o}wen}},
  \bibinfo{author}{\bibfnamefont{C.}~\bibnamefont{Bechinger}},
  \bibnamefont{and} \bibinfo{author}{\bibfnamefont{T.}~\bibnamefont{Speck}},
  \bibinfo{journal}{Phys. Rev. Lett.} \textbf{\bibinfo{volume}{110}},
  \bibinfo{pages}{238301} (\bibinfo{year}{2013}).

\bibitem[{\citenamefont{Speck et~al.}(2014)\citenamefont{Speck, Bialk{\'e},
  Menzel, and L{\"o}wen}}]{speck2014effective}
\bibinfo{author}{\bibfnamefont{T.}~\bibnamefont{Speck}},
  \bibinfo{author}{\bibfnamefont{J.}~\bibnamefont{Bialk{\'e}}},
  \bibinfo{author}{\bibfnamefont{A.~M.} \bibnamefont{Menzel}},
  \bibnamefont{and}
  \bibinfo{author}{\bibfnamefont{H.}~\bibnamefont{L{\"o}wen}},
  \bibinfo{journal}{Phys. Rev. Lett.} \textbf{\bibinfo{volume}{112}},
  \bibinfo{pages}{218304} (\bibinfo{year}{2014}).

\bibitem[{\citenamefont{Speck et~al.}(2015)\citenamefont{Speck, Menzel,
  Bialk{\'e}, and L{\"o}wen}}]{speck2015dynamical}
\bibinfo{author}{\bibfnamefont{T.}~\bibnamefont{Speck}},
  \bibinfo{author}{\bibfnamefont{A.~M.} \bibnamefont{Menzel}},
  \bibinfo{author}{\bibfnamefont{J.}~\bibnamefont{Bialk{\'e}}},
  \bibnamefont{and}
  \bibinfo{author}{\bibfnamefont{H.}~\bibnamefont{L{\"o}wen}},
  \bibinfo{journal}{J. Chem. Phys.} \textbf{\bibinfo{volume}{142}},
  \bibinfo{pages}{224109} (\bibinfo{year}{2015}).

\bibitem[{\citenamefont{Bialk{\'e} et~al.}(2012)\citenamefont{Bialk{\'e},
  Speck, and L{\"o}wen}}]{bialke2012crystallization}
\bibinfo{author}{\bibfnamefont{J.}~\bibnamefont{Bialk{\'e}}},
  \bibinfo{author}{\bibfnamefont{T.}~\bibnamefont{Speck}}, \bibnamefont{and}
  \bibinfo{author}{\bibfnamefont{H.}~\bibnamefont{L{\"o}wen}},
  \bibinfo{journal}{Phys. Rev. Lett.} \textbf{\bibinfo{volume}{108}},
  \bibinfo{pages}{168301} (\bibinfo{year}{2012}).

\bibitem[{\citenamefont{Menzel and L{\"o}wen}(2013)}]{menzel2013traveling}
\bibinfo{author}{\bibfnamefont{A.~M.} \bibnamefont{Menzel}} \bibnamefont{and}
  \bibinfo{author}{\bibfnamefont{H.}~\bibnamefont{L{\"o}wen}},
  \bibinfo{journal}{Phys. Rev. Lett.} \textbf{\bibinfo{volume}{110}},
  \bibinfo{pages}{055702} (\bibinfo{year}{2013}).

\bibitem[{\citenamefont{Menzel et~al.}(2014)\citenamefont{Menzel, Ohta, and
  L{\"o}wen}}]{menzel2014active}
\bibinfo{author}{\bibfnamefont{A.~M.} \bibnamefont{Menzel}},
  \bibinfo{author}{\bibfnamefont{T.}~\bibnamefont{Ohta}}, \bibnamefont{and}
  \bibinfo{author}{\bibfnamefont{H.}~\bibnamefont{L{\"o}wen}},
  \bibinfo{journal}{Phys. Rev. E} \textbf{\bibinfo{volume}{89}},
  \bibinfo{pages}{022301} (\bibinfo{year}{2014}).

\bibitem[{\citenamefont{Wensink and L{\"o}wen}(2012)}]{wensink2012emergent}
\bibinfo{author}{\bibfnamefont{H.~H.} \bibnamefont{Wensink}} \bibnamefont{and}
  \bibinfo{author}{\bibfnamefont{H.}~\bibnamefont{L{\"o}wen}},
  \bibinfo{journal}{J. Phys.: Condens. Matter} \textbf{\bibinfo{volume}{24}},
  \bibinfo{pages}{464130} (\bibinfo{year}{2012}).

\bibitem[{\citenamefont{Menzel and Ohta}(2012)}]{menzel2012soft}
\bibinfo{author}{\bibfnamefont{A.~M.} \bibnamefont{Menzel}} \bibnamefont{and}
  \bibinfo{author}{\bibfnamefont{T.}~\bibnamefont{Ohta}}, \bibinfo{journal}{EPL
  (Europhys. Lett.)} \textbf{\bibinfo{volume}{99}}, \bibinfo{pages}{58001}
  (\bibinfo{year}{2012}).

\bibitem[{\citenamefont{McCandlish et~al.}(2012)\citenamefont{McCandlish,
  Baskaran, and Hagan}}]{mccandlish2012spontaneous}
\bibinfo{author}{\bibfnamefont{S.~R.} \bibnamefont{McCandlish}},
  \bibinfo{author}{\bibfnamefont{A.}~\bibnamefont{Baskaran}}, \bibnamefont{and}
  \bibinfo{author}{\bibfnamefont{M.~F.} \bibnamefont{Hagan}},
  \bibinfo{journal}{Soft Matter} \textbf{\bibinfo{volume}{8}},
  \bibinfo{pages}{2527} (\bibinfo{year}{2012}).

\bibitem[{\citenamefont{Menzel}(2013)}]{menzel2013unidirectional}
\bibinfo{author}{\bibfnamefont{A.~M.} \bibnamefont{Menzel}},
  \bibinfo{journal}{J. Phys.: Condens. Matter} \textbf{\bibinfo{volume}{25}},
  \bibinfo{pages}{505103} (\bibinfo{year}{2013}).

\bibitem[{\citenamefont{Kogler and Klapp}(2015)}]{kogler2015lane}
\bibinfo{author}{\bibfnamefont{F.}~\bibnamefont{Kogler}} \bibnamefont{and}
  \bibinfo{author}{\bibfnamefont{S.~H.~L.} \bibnamefont{Klapp}},
  \bibinfo{journal}{EPL (Europhys. Lett.)} \textbf{\bibinfo{volume}{110}},
  \bibinfo{pages}{10004} (\bibinfo{year}{2015}).

\bibitem[{\citenamefont{Ballerini et~al.}(2008)\citenamefont{Ballerini,
  Cabibbo, Candelier, Cavagna, Cisbani, Giardina, Lecomte, Orlandi, Parisi,
  Procaccini et~al.}}]{ballerini2008interaction}
\bibinfo{author}{\bibfnamefont{M.}~\bibnamefont{Ballerini}},
  \bibinfo{author}{\bibfnamefont{N.}~\bibnamefont{Cabibbo}},
  \bibinfo{author}{\bibfnamefont{R.}~\bibnamefont{Candelier}},
  \bibinfo{author}{\bibfnamefont{A.}~\bibnamefont{Cavagna}},
  \bibinfo{author}{\bibfnamefont{E.}~\bibnamefont{Cisbani}},
  \bibinfo{author}{\bibfnamefont{I.}~\bibnamefont{Giardina}},
  \bibinfo{author}{\bibfnamefont{V.}~\bibnamefont{Lecomte}},
  \bibinfo{author}{\bibfnamefont{A.}~\bibnamefont{Orlandi}},
  \bibinfo{author}{\bibfnamefont{G.}~\bibnamefont{Parisi}},
  \bibinfo{author}{\bibfnamefont{A.}~\bibnamefont{Procaccini}},
  \bibnamefont{et~al.}, \bibinfo{journal}{Proc. Natl. Acad. Sci. USA}
  \textbf{\bibinfo{volume}{105}}, \bibinfo{pages}{1232} (\bibinfo{year}{2008}).

\bibitem[{\citenamefont{Sokolov et~al.}(2007)\citenamefont{Sokolov, Aranson,
  Kessler, and Goldstein}}]{sokolov2007concentration}
\bibinfo{author}{\bibfnamefont{A.}~\bibnamefont{Sokolov}},
  \bibinfo{author}{\bibfnamefont{I.~S.} \bibnamefont{Aranson}},
  \bibinfo{author}{\bibfnamefont{J.~O.} \bibnamefont{Kessler}},
  \bibnamefont{and} \bibinfo{author}{\bibfnamefont{R.~E.}
  \bibnamefont{Goldstein}}, \bibinfo{journal}{Phys. Rev. Lett.}
  \textbf{\bibinfo{volume}{98}}, \bibinfo{pages}{158102}
  (\bibinfo{year}{2007}).

\bibitem[{\citenamefont{Aranson and Tsimring}(2006)}]{aranson2006patterns}
\bibinfo{author}{\bibfnamefont{I.~S.} \bibnamefont{Aranson}} \bibnamefont{and}
  \bibinfo{author}{\bibfnamefont{L.~S.} \bibnamefont{Tsimring}},
  \bibinfo{journal}{Rev. Mod. Phys.} \textbf{\bibinfo{volume}{78}},
  \bibinfo{pages}{641} (\bibinfo{year}{2006}).

\bibitem[{\citenamefont{Deseigne et~al.}(2010)\citenamefont{Deseigne, Dauchot,
  and Chat{\'e}}}]{deseigne2010collective}
\bibinfo{author}{\bibfnamefont{J.}~\bibnamefont{Deseigne}},
  \bibinfo{author}{\bibfnamefont{O.}~\bibnamefont{Dauchot}}, \bibnamefont{and}
  \bibinfo{author}{\bibfnamefont{H.}~\bibnamefont{Chat{\'e}}},
  \bibinfo{journal}{Phys. Rev. Lett.} \textbf{\bibinfo{volume}{105}},
  \bibinfo{pages}{098001} (\bibinfo{year}{2010}).

\bibitem[{\citenamefont{Kudrolli et~al.}(2008)\citenamefont{Kudrolli, Lumay,
  Volfson, and Tsimring}}]{kudrolli2008swarming}
\bibinfo{author}{\bibfnamefont{A.}~\bibnamefont{Kudrolli}},
  \bibinfo{author}{\bibfnamefont{G.}~\bibnamefont{Lumay}},
  \bibinfo{author}{\bibfnamefont{D.}~\bibnamefont{Volfson}}, \bibnamefont{and}
  \bibinfo{author}{\bibfnamefont{L.~S.} \bibnamefont{Tsimring}},
  \bibinfo{journal}{Phys. Rev. Lett.} \textbf{\bibinfo{volume}{100}},
  \bibinfo{pages}{058001} (\bibinfo{year}{2008}).

\bibitem[{\citenamefont{Paxton et~al.}(2004)\citenamefont{Paxton, Kistler,
  Olmeda, Sen, St.~Angelo, Cao, Mallouk, Lammert, and
  Crespi}}]{paxton2004catalytic}
\bibinfo{author}{\bibfnamefont{W.~F.} \bibnamefont{Paxton}},
  \bibinfo{author}{\bibfnamefont{K.~C.} \bibnamefont{Kistler}},
  \bibinfo{author}{\bibfnamefont{C.~C.} \bibnamefont{Olmeda}},
  \bibinfo{author}{\bibfnamefont{A.}~\bibnamefont{Sen}},
  \bibinfo{author}{\bibfnamefont{S.~K.} \bibnamefont{St.~Angelo}},
  \bibinfo{author}{\bibfnamefont{Y.}~\bibnamefont{Cao}},
  \bibinfo{author}{\bibfnamefont{T.~E.} \bibnamefont{Mallouk}},
  \bibinfo{author}{\bibfnamefont{P.~E.} \bibnamefont{Lammert}},
  \bibnamefont{and} \bibinfo{author}{\bibfnamefont{V.~H.}
  \bibnamefont{Crespi}}, \bibinfo{journal}{J. Am. Chem. Soc.}
  \textbf{\bibinfo{volume}{126}}, \bibinfo{pages}{13424}
  (\bibinfo{year}{2004}).

\bibitem[{\citenamefont{Tierno et~al.}(2010)\citenamefont{Tierno, Albalat, and
  Sagu{\'e}s}}]{tierno2010autonomously}
\bibinfo{author}{\bibfnamefont{P.}~\bibnamefont{Tierno}},
  \bibinfo{author}{\bibfnamefont{R.}~\bibnamefont{Albalat}}, \bibnamefont{and}
  \bibinfo{author}{\bibfnamefont{F.}~\bibnamefont{Sagu{\'e}s}},
  \bibinfo{journal}{Small} \textbf{\bibinfo{volume}{6}}, \bibinfo{pages}{1749}
  (\bibinfo{year}{2010}).

\bibitem[{\citenamefont{ten Hagen et~al.}(2015)\citenamefont{ten Hagen,
  Wittkowski, Takagi, K{\"u}mmel, Bechinger, and L{\"o}wen}}]{hagen2014can}
\bibinfo{author}{\bibfnamefont{B.}~\bibnamefont{ten Hagen}},
  \bibinfo{author}{\bibfnamefont{R.}~\bibnamefont{Wittkowski}},
  \bibinfo{author}{\bibfnamefont{D.}~\bibnamefont{Takagi}},
  \bibinfo{author}{\bibfnamefont{F.}~\bibnamefont{K{\"u}mmel}},
  \bibinfo{author}{\bibfnamefont{C.}~\bibnamefont{Bechinger}},
  \bibnamefont{and}
  \bibinfo{author}{\bibfnamefont{H.}~\bibnamefont{L{\"o}wen}},
  \bibinfo{journal}{J. Phys.: Condens. Matter} \textbf{\bibinfo{volume}{27}},
  \bibinfo{pages}{194110} (\bibinfo{year}{2015}).

\bibitem[{\citenamefont{Baskaran and Marchetti}(2008)}]{baskaran2008enhanced}
\bibinfo{author}{\bibfnamefont{A.}~\bibnamefont{Baskaran}} \bibnamefont{and}
  \bibinfo{author}{\bibfnamefont{M.~C.} \bibnamefont{Marchetti}},
  \bibinfo{journal}{Phys. Rev. Lett.} \textbf{\bibinfo{volume}{101}},
  \bibinfo{pages}{268101} (\bibinfo{year}{2008}).

\bibitem[{\citenamefont{Baskaran and
  Marchetti}(2009)}]{baskaran2009statistical}
\bibinfo{author}{\bibfnamefont{A.}~\bibnamefont{Baskaran}} \bibnamefont{and}
  \bibinfo{author}{\bibfnamefont{M.~C.} \bibnamefont{Marchetti}},
  \bibinfo{journal}{Proc. Natl. Acad. Sci. USA} \textbf{\bibinfo{volume}{106}},
  \bibinfo{pages}{15567} (\bibinfo{year}{2009}).

\bibitem[{\citenamefont{Chou et~al.}(2012)\citenamefont{Chou, Wolfe, and
  Ihle}}]{chou2012kinetic}
\bibinfo{author}{\bibfnamefont{Y.-L.} \bibnamefont{Chou}},
  \bibinfo{author}{\bibfnamefont{R.}~\bibnamefont{Wolfe}}, \bibnamefont{and}
  \bibinfo{author}{\bibfnamefont{T.}~\bibnamefont{Ihle}},
  \bibinfo{journal}{Phys. Rev. E} \textbf{\bibinfo{volume}{86}},
  \bibinfo{pages}{012120} (\bibinfo{year}{2012}).

\bibitem[{\citenamefont{Grossmann et~al.}(2015)\citenamefont{Grossmann,
  Romanczuk, B{\"a}r, and Schimansky-Geier}}]{grossmann2015pattern}
\bibinfo{author}{\bibfnamefont{R.}~\bibnamefont{Grossmann}},
  \bibinfo{author}{\bibfnamefont{P.}~\bibnamefont{Romanczuk}},
  \bibinfo{author}{\bibfnamefont{M.}~\bibnamefont{B{\"a}r}}, \bibnamefont{and}
  \bibinfo{author}{\bibfnamefont{L.}~\bibnamefont{Schimansky-Geier}},
  \bibinfo{journal}{Eur. Phys. J. Special Topics}
  \textbf{\bibinfo{volume}{224}}, \bibinfo{pages}{1325} (\bibinfo{year}{2015}).

\bibitem[{\citenamefont{Heidenreich et~al.}(2015)\citenamefont{Heidenreich,
  Dunkel, Klapp, and B{\"a}r}}]{heidenreich2015hydrodynamic}
\bibinfo{author}{\bibfnamefont{S.}~\bibnamefont{Heidenreich}},
  \bibinfo{author}{\bibfnamefont{J.}~\bibnamefont{Dunkel}},
  \bibinfo{author}{\bibfnamefont{S.~H.~L.} \bibnamefont{Klapp}},
  \bibnamefont{and} \bibinfo{author}{\bibfnamefont{M.}~\bibnamefont{B{\"a}r}},
  \bibinfo{journal}{arXiv preprint arXiv:1509.08661}  (\bibinfo{year}{2015}).

\bibitem[{\citenamefont{Chou and Ihle}(2015)}]{chou2015active}
\bibinfo{author}{\bibfnamefont{Y.-L.} \bibnamefont{Chou}} \bibnamefont{and}
  \bibinfo{author}{\bibfnamefont{T.}~\bibnamefont{Ihle}},
  \bibinfo{journal}{Phys. Rev. E} \textbf{\bibinfo{volume}{91}},
  \bibinfo{pages}{022103} (\bibinfo{year}{2015}).

\bibitem[{\citenamefont{Aranson et~al.}(2007)\citenamefont{Aranson, Sokolov,
  Kessler, and Goldstein}}]{aranson2007model}
\bibinfo{author}{\bibfnamefont{I.~S.} \bibnamefont{Aranson}},
  \bibinfo{author}{\bibfnamefont{A.}~\bibnamefont{Sokolov}},
  \bibinfo{author}{\bibfnamefont{J.~O.} \bibnamefont{Kessler}},
  \bibnamefont{and} \bibinfo{author}{\bibfnamefont{R.~E.}
  \bibnamefont{Goldstein}}, \bibinfo{journal}{Phys. Rev. E}
  \textbf{\bibinfo{volume}{75}}, \bibinfo{pages}{040901}
  (\bibinfo{year}{2007}).

\bibitem[{\citenamefont{Pleiner et~al.}(2013)\citenamefont{Pleiner,
  Sven{\v{s}}ek, and Brand}}]{pleiner2013active}
\bibinfo{author}{\bibfnamefont{H.}~\bibnamefont{Pleiner}},
  \bibinfo{author}{\bibfnamefont{D.}~\bibnamefont{Sven{\v{s}}ek}},
  \bibnamefont{and} \bibinfo{author}{\bibfnamefont{H.~R.} \bibnamefont{Brand}},
  \bibinfo{journal}{Eur. Phys. J. E} \textbf{\bibinfo{volume}{36}},
  \bibinfo{pages}{1} (\bibinfo{year}{2013}).

\bibitem[{\citenamefont{Brand et~al.}(2014)\citenamefont{Brand, Pleiner, and
  Sven{\v{s}}ek}}]{brand2014reversible}
\bibinfo{author}{\bibfnamefont{H.~R.} \bibnamefont{Brand}},
  \bibinfo{author}{\bibfnamefont{H.}~\bibnamefont{Pleiner}}, \bibnamefont{and}
  \bibinfo{author}{\bibfnamefont{D.}~\bibnamefont{Sven{\v{s}}ek}},
  \bibinfo{journal}{Eur. Phys. J. E} \textbf{\bibinfo{volume}{37}},
  \bibinfo{pages}{1} (\bibinfo{year}{2014}).

\bibitem[{\citenamefont{Marconi and Tarazona}(1999)}]{marconi1999dynamic}
\bibinfo{author}{\bibfnamefont{U.~M.~B.} \bibnamefont{Marconi}}
  \bibnamefont{and} \bibinfo{author}{\bibfnamefont{P.}~\bibnamefont{Tarazona}},
  \bibinfo{journal}{J. Chem. Phys.} \textbf{\bibinfo{volume}{110}},
  \bibinfo{pages}{8032} (\bibinfo{year}{1999}).

\bibitem[{\citenamefont{Marconi and Tarazona}(2000)}]{marconi2000dynamic}
\bibinfo{author}{\bibfnamefont{U.~M.~B.} \bibnamefont{Marconi}}
  \bibnamefont{and} \bibinfo{author}{\bibfnamefont{P.}~\bibnamefont{Tarazona}},
  \bibinfo{journal}{J. Phys.: Condens. Matter} \textbf{\bibinfo{volume}{12}},
  \bibinfo{pages}{A413} (\bibinfo{year}{2000}).

\bibitem[{\citenamefont{Archer and Evans}(2004)}]{archer2004dynamical}
\bibinfo{author}{\bibfnamefont{A.~J.} \bibnamefont{Archer}} \bibnamefont{and}
  \bibinfo{author}{\bibfnamefont{R.}~\bibnamefont{Evans}}, \bibinfo{journal}{J.
  Chem. Phys.} \textbf{\bibinfo{volume}{121}}, \bibinfo{pages}{4246}
  (\bibinfo{year}{2004}).

\bibitem[{\citenamefont{Archer}(2005)}]{archer2005dynamical}
\bibinfo{author}{\bibfnamefont{A.\hspace{0.07cm}J.}\hspace{0.07cm}\bibnamefont{Archer}},\hspace{0.08cm}\bibinfo{journal}{J.\hspace{0.07cm}Phys.:\hspace{0.07cm}Condens.\hspace{0.07cm}Matter}\hspace{0.07cm}\textbf{\bibinfo{volume}{17}},\hspace{0.07cm}\bibinfo{pages}{1405}\hspace{0.07cm}(\bibinfo{year}{2005}).

\bibitem[{\citenamefont{van Teeffelen et~al.}(2008)\citenamefont{van Teeffelen,
  Likos, and L{\"o}wen}}]{van2008colloidal}
\bibinfo{author}{\bibfnamefont{S.}~\bibnamefont{van Teeffelen}},
  \bibinfo{author}{\bibfnamefont{C.~N.} \bibnamefont{Likos}}, \bibnamefont{and}
  \bibinfo{author}{\bibfnamefont{H.}~\bibnamefont{L{\"o}wen}},
  \bibinfo{journal}{Phys. Rev. Lett.} \textbf{\bibinfo{volume}{100}},
  \bibinfo{pages}{108302} (\bibinfo{year}{2008}).

\bibitem[{\citenamefont{Penna et~al.}(2003)\citenamefont{Penna, Dzubiella, and
  Tarazona}}]{penna2003dynamic}
\bibinfo{author}{\bibfnamefont{F.}~\bibnamefont{Penna}},
  \bibinfo{author}{\bibfnamefont{J.}~\bibnamefont{Dzubiella}},
  \bibnamefont{and} \bibinfo{author}{\bibfnamefont{P.}~\bibnamefont{Tarazona}},
  \bibinfo{journal}{Phys. Rev. E} \textbf{\bibinfo{volume}{68}},
  \bibinfo{pages}{061407} (\bibinfo{year}{2003}).

\bibitem[{\citenamefont{Wittkowski et~al.}(2012)\citenamefont{Wittkowski,
  L{\"o}wen, and Brand}}]{wittkowski2012extended}
\bibinfo{author}{\bibfnamefont{R.}~\bibnamefont{Wittkowski}},
  \bibinfo{author}{\bibfnamefont{H.}~\bibnamefont{L{\"o}wen}},
  \bibnamefont{and} \bibinfo{author}{\bibfnamefont{H.~R.} \bibnamefont{Brand}},
  \bibinfo{journal}{J. Chem. Phys.} \textbf{\bibinfo{volume}{137}},
  \bibinfo{pages}{224904} (\bibinfo{year}{2012}).

\bibitem[{\citenamefont{Archer et~al.}(2010)\citenamefont{Archer, Robbins, and
  Thiele}}]{archer2010dynamical}
\bibinfo{author}{\bibfnamefont{A.~J.} \bibnamefont{Archer}},
  \bibinfo{author}{\bibfnamefont{M.~J.} \bibnamefont{Robbins}},
  \bibnamefont{and} \bibinfo{author}{\bibfnamefont{U.}~\bibnamefont{Thiele}},
  \bibinfo{journal}{Phys. Rev. E} \textbf{\bibinfo{volume}{81}},
  \bibinfo{pages}{021602} (\bibinfo{year}{2010}).

\bibitem[{\citenamefont{Wittkowski et~al.}(2010)\citenamefont{Wittkowski,
  L{\"o}wen, and Brand}}]{wittkowski2010derivation}
\bibinfo{author}{\bibfnamefont{R.}~\bibnamefont{Wittkowski}},
  \bibinfo{author}{\bibfnamefont{H.}~\bibnamefont{L{\"o}wen}},
  \bibnamefont{and} \bibinfo{author}{\bibfnamefont{H.~R.} \bibnamefont{Brand}},
  \bibinfo{journal}{Phys. Rev. E} \textbf{\bibinfo{volume}{82}},
  \bibinfo{pages}{031708} (\bibinfo{year}{2010}).

\bibitem[{\citenamefont{Aerov and Kr{\"u}ger}(2014)}]{aerov2014driven}
\bibinfo{author}{\bibfnamefont{A.~A.} \bibnamefont{Aerov}} \bibnamefont{and}
  \bibinfo{author}{\bibfnamefont{M.}~\bibnamefont{Kr{\"u}ger}},
  \bibinfo{journal}{J. Chem. Phys.} \textbf{\bibinfo{volume}{140}},
  \bibinfo{pages}{094701} (\bibinfo{year}{2014}).

\bibitem[{\citenamefont{Aerov and Kr{\"u}ger}(2015)}]{aerov2015theory}
\bibinfo{author}{\bibfnamefont{A.~A.} \bibnamefont{Aerov}} \bibnamefont{and}
  \bibinfo{author}{\bibfnamefont{M.}~\bibnamefont{Kr{\"u}ger}},
  \bibinfo{journal}{Phys. Rev. E} \textbf{\bibinfo{volume}{92}},
  \bibinfo{pages}{042301} (\bibinfo{year}{2015}).

\bibitem[{\citenamefont{Rex and L{\"o}wen}(2008)}]{rex2008dynamical}
\bibinfo{author}{\bibfnamefont{M.}~\bibnamefont{Rex}} \bibnamefont{and}
  \bibinfo{author}{\bibfnamefont{H.}~\bibnamefont{L{\"o}wen}},
  \bibinfo{journal}{Phys. Rev. Lett.} \textbf{\bibinfo{volume}{101}},
  \bibinfo{pages}{148302} (\bibinfo{year}{2008}).

\bibitem[{\citenamefont{Rex and L{\"o}wen}(2009)}]{rex2009dynamical}
\bibinfo{author}{\bibfnamefont{M.}~\bibnamefont{Rex}} \bibnamefont{and}
  \bibinfo{author}{\bibfnamefont{H.}~\bibnamefont{L{\"o}wen}},
  \bibinfo{journal}{Eur. Phys. J. E} \textbf{\bibinfo{volume}{28}},
  \bibinfo{pages}{139} (\bibinfo{year}{2009}).

\bibitem[{\citenamefont{Donev and Vanden-Eijnden}(2014)}]{donev2014dynamic}
\bibinfo{author}{\bibfnamefont{A.}~\bibnamefont{Donev}} \bibnamefont{and}
  \bibinfo{author}{\bibfnamefont{E.}~\bibnamefont{Vanden-Eijnden}},
  \bibinfo{journal}{J. Chem. Phys.} \textbf{\bibinfo{volume}{140}},
  \bibinfo{pages}{234115} (\bibinfo{year}{2014}).

\bibitem[{\citenamefont{Wensink and L{\"o}wen}(2008)}]{wensink2008aggregation}
\bibinfo{author}{\bibfnamefont{H.~H.} \bibnamefont{Wensink}} \bibnamefont{and}
  \bibinfo{author}{\bibfnamefont{H.}~\bibnamefont{L{\"o}wen}},
  \bibinfo{journal}{Phys. Rev. E} \textbf{\bibinfo{volume}{78}},
  \bibinfo{pages}{031409} (\bibinfo{year}{2008}).

\bibitem[{\citenamefont{Wittkowski and
  L{\"o}wen}(2011)}]{wittkowski2011dynamical}
\bibinfo{author}{\bibfnamefont{R.}~\bibnamefont{Wittkowski}} \bibnamefont{and}
  \bibinfo{author}{\bibfnamefont{H.}~\bibnamefont{L{\"o}wen}},
  \bibinfo{journal}{Mol. Phys.} \textbf{\bibinfo{volume}{109}},
  \bibinfo{pages}{2935} (\bibinfo{year}{2011}).

\bibitem[{\citenamefont{Nash et~al.}(2010)\citenamefont{Nash, Adhikari,
  Tailleur, and Cates}}]{nash2010run}
\bibinfo{author}{\bibfnamefont{R.~W.} \bibnamefont{Nash}},
  \bibinfo{author}{\bibfnamefont{R.}~\bibnamefont{Adhikari}},
  \bibinfo{author}{\bibfnamefont{J.}~\bibnamefont{Tailleur}}, \bibnamefont{and}
  \bibinfo{author}{\bibfnamefont{M.~E.} \bibnamefont{Cates}},
  \bibinfo{journal}{Phys. Rev. Lett.} \textbf{\bibinfo{volume}{104}},
  \bibinfo{pages}{258101} (\bibinfo{year}{2010}).

\bibitem[{\citenamefont{Hennes et~al.}(2014)\citenamefont{Hennes, Wolff, and
  Stark}}]{hennes2014self}
\bibinfo{author}{\bibfnamefont{M.}~\bibnamefont{Hennes}},
  \bibinfo{author}{\bibfnamefont{K.}~\bibnamefont{Wolff}}, \bibnamefont{and}
  \bibinfo{author}{\bibfnamefont{H.}~\bibnamefont{Stark}},
  \bibinfo{journal}{Phys. Rev. Lett.} \textbf{\bibinfo{volume}{112}},
  \bibinfo{pages}{238104} (\bibinfo{year}{2014}).

\bibitem[{\citenamefont{Purcell}(1977)}]{purcell1977life}
\bibinfo{author}{\bibfnamefont{E.~M.} \bibnamefont{Purcell}},
  \bibinfo{journal}{Am. J. Phys.} \textbf{\bibinfo{volume}{45}},
  \bibinfo{pages}{3} (\bibinfo{year}{1977}).

\bibitem[{\citenamefont{Aditi~Simha and
  Ramaswamy}(2002)}]{aditi2002hydrodynamic}
\bibinfo{author}{\bibfnamefont{R.}~\bibnamefont{Aditi~Simha}} \bibnamefont{and}
  \bibinfo{author}{\bibfnamefont{S.}~\bibnamefont{Ramaswamy}},
  \bibinfo{journal}{Phys. Rev. Lett.} \textbf{\bibinfo{volume}{89}},
  \bibinfo{pages}{058101} (\bibinfo{year}{2002}).

\bibitem[{\citenamefont{Hatwalne et~al.}(2004)\citenamefont{Hatwalne,
  Ramaswamy, Rao, and Simha}}]{hatwalne2004rheology}
\bibinfo{author}{\bibfnamefont{Y.}~\bibnamefont{Hatwalne}},
  \bibinfo{author}{\bibfnamefont{S.}~\bibnamefont{Ramaswamy}},
  \bibinfo{author}{\bibfnamefont{M.}~\bibnamefont{Rao}}, \bibnamefont{and}
  \bibinfo{author}{\bibfnamefont{R.~A.} \bibnamefont{Simha}},
  \bibinfo{journal}{Phys. Rev. Lett.} \textbf{\bibinfo{volume}{92}},
  \bibinfo{pages}{118101} (\bibinfo{year}{2004}).

\bibitem[{\citenamefont{Golestanian}(2008)}]{golestanian2008three}
\bibinfo{author}{\bibfnamefont{R.}~\bibnamefont{Golestanian}},
  \bibinfo{journal}{Eur. Phys. J. E} \textbf{\bibinfo{volume}{25}},
  \bibinfo{pages}{1} (\bibinfo{year}{2008}).

\bibitem[{\citenamefont{Fily et~al.}(2012)\citenamefont{Fily, Baskaran, and
  Marchetti}}]{fily2012cooperative}
\bibinfo{author}{\bibfnamefont{Y.}~\bibnamefont{Fily}},
  \bibinfo{author}{\bibfnamefont{A.}~\bibnamefont{Baskaran}}, \bibnamefont{and}
  \bibinfo{author}{\bibfnamefont{M.~C.} \bibnamefont{Marchetti}},
  \bibinfo{journal}{Soft Matter} \textbf{\bibinfo{volume}{8}},
  \bibinfo{pages}{3002} (\bibinfo{year}{2012}).

\bibitem[{\citenamefont{Happel and Brenner}(1983)}]{happel2012low}
\bibinfo{author}{\bibfnamefont{J.}~\bibnamefont{Happel}} \bibnamefont{and}
  \bibinfo{author}{\bibfnamefont{H.}~\bibnamefont{Brenner}},
  \emph{\bibinfo{title}{Low Reynolds Number Hydrodynamics}}
  (\bibinfo{publisher}{Martinus Neijhoff Publishers}, \bibinfo{year}{1983}).

\bibitem[{\citenamefont{Dhont}(1996)}]{dhont1996introduction}
\bibinfo{author}{\bibfnamefont{J.~K.~G.} \bibnamefont{Dhont}},
  \emph{\bibinfo{title}{An Introduction to Dynamics of Colloids}}
  (\bibinfo{publisher}{Elsevier}, \bibinfo{year}{1996}).

\bibitem[{\citenamefont{Reichert and Stark}(2004)}]{reichert2004hydrodynamic}
\bibinfo{author}{\bibfnamefont{M.}~\bibnamefont{Reichert}} \bibnamefont{and}
  \bibinfo{author}{\bibfnamefont{H.}~\bibnamefont{Stark}},
  \bibinfo{journal}{Phys. Rev. E} \textbf{\bibinfo{volume}{69}},
  \bibinfo{pages}{031407} (\bibinfo{year}{2004}).

\bibitem[{\citenamefont{Brenner}(1963)}]{brenner1963stokes}
\bibinfo{author}{\bibfnamefont{H.}~\bibnamefont{Brenner}},
  \bibinfo{journal}{Chem. Eng. Sci.} \textbf{\bibinfo{volume}{18}},
  \bibinfo{pages}{1} (\bibinfo{year}{1963}).

\bibitem[{\citenamefont{Mazur and Van~Saarloos}(1982)}]{mazur1982many}
\bibinfo{author}{\bibfnamefont{P.}~\bibnamefont{Mazur}} \bibnamefont{and}
  \bibinfo{author}{\bibfnamefont{W.}~\bibnamefont{Van~Saarloos}},
  \bibinfo{journal}{Physica A} \textbf{\bibinfo{volume}{115}},
  \bibinfo{pages}{21} (\bibinfo{year}{1982}).

\bibitem[{\citenamefont{Evans et~al.}(2011)\citenamefont{Evans, Ishikawa,
  Yamaguchi, and Lauga}}]{evans2011orientational}
\bibinfo{author}{\bibfnamefont{A.~A.} \bibnamefont{Evans}},
  \bibinfo{author}{\bibfnamefont{T.}~\bibnamefont{Ishikawa}},
  \bibinfo{author}{\bibfnamefont{T.}~\bibnamefont{Yamaguchi}},
  \bibnamefont{and} \bibinfo{author}{\bibfnamefont{E.}~\bibnamefont{Lauga}},
  \bibinfo{journal}{Phys. Fluids} \textbf{\bibinfo{volume}{23}},
  \bibinfo{pages}{111702} (\bibinfo{year}{2011}).

\bibitem[{\citenamefont{Alarcon and
  Pagonabarraga}(2013)}]{alarcon2013spontaneous}
\bibinfo{author}{\bibfnamefont{F.}~\bibnamefont{Alarcon}} \bibnamefont{and}
  \bibinfo{author}{\bibfnamefont{I.}~\bibnamefont{Pagonabarraga}},
  \bibinfo{journal}{J. Mol. Liq.} \textbf{\bibinfo{volume}{185}},
  \bibinfo{pages}{56} (\bibinfo{year}{2013}).

\bibitem[{\citenamefont{Z{\"o}ttl and Stark}(2014)}]{zottl2014hydrodynamics}
\bibinfo{author}{\bibfnamefont{A.}~\bibnamefont{Z{\"o}ttl}} \bibnamefont{and}
  \bibinfo{author}{\bibfnamefont{H.}~\bibnamefont{Stark}},
  \bibinfo{journal}{Phys. Rev. Lett.} \textbf{\bibinfo{volume}{112}},
  \bibinfo{pages}{118101} (\bibinfo{year}{2014}).

\bibitem[{\citenamefont{Navarro and
  Pagonabarraga}(2010)}]{navarro2010hydrodynamic}
\bibinfo{author}{\bibfnamefont{R.~M.} \bibnamefont{Navarro}} \bibnamefont{and}
  \bibinfo{author}{\bibfnamefont{I.}~\bibnamefont{Pagonabarraga}},
  \bibinfo{journal}{Eur. Phys. J. E} \textbf{\bibinfo{volume}{33}},
  \bibinfo{pages}{27} (\bibinfo{year}{2010}).

\bibitem[{\citenamefont{Mladek et~al.}(2006)\citenamefont{Mladek, Gottwald,
  Kahl, Neumann, and Likos}}]{mladek2006formation}
\bibinfo{author}{\bibfnamefont{B.~M.} \bibnamefont{Mladek}},
  \bibinfo{author}{\bibfnamefont{D.}~\bibnamefont{Gottwald}},
  \bibinfo{author}{\bibfnamefont{G.}~\bibnamefont{Kahl}},
  \bibinfo{author}{\bibfnamefont{M.}~\bibnamefont{Neumann}}, \bibnamefont{and}
  \bibinfo{author}{\bibfnamefont{C.~N.} \bibnamefont{Likos}},
  \bibinfo{journal}{Phys. Rev. Lett.} \textbf{\bibinfo{volume}{96}},
  \bibinfo{pages}{045701} (\bibinfo{year}{2006}).

\bibitem[{\citenamefont{Wensink et~al.}(2014)\citenamefont{Wensink, Kantsler,
  Goldstein, and Dunkel}}]{wensink2014controlling}
\bibinfo{author}{\bibfnamefont{H.~H.} \bibnamefont{Wensink}},
  \bibinfo{author}{\bibfnamefont{V.}~\bibnamefont{Kantsler}},
  \bibinfo{author}{\bibfnamefont{R.~E.} \bibnamefont{Goldstein}},
  \bibnamefont{and} \bibinfo{author}{\bibfnamefont{J.}~\bibnamefont{Dunkel}},
  \bibinfo{journal}{Phys. Rev. E} \textbf{\bibinfo{volume}{89}},
  \bibinfo{pages}{010302} (\bibinfo{year}{2014}).

\bibitem[{\citenamefont{Archer et~al.}(2014)\citenamefont{Archer, Walters,
  Thiele, and Knobloch}}]{archer2014solidification}
\bibinfo{author}{\bibfnamefont{A.~J.} \bibnamefont{Archer}},
  \bibinfo{author}{\bibfnamefont{M.~C.} \bibnamefont{Walters}},
  \bibinfo{author}{\bibfnamefont{U.}~\bibnamefont{Thiele}}, \bibnamefont{and}
  \bibinfo{author}{\bibfnamefont{E.}~\bibnamefont{Knobloch}},
  \bibinfo{journal}{Phys. Rev. E} \textbf{\bibinfo{volume}{90}},
  \bibinfo{pages}{042404} (\bibinfo{year}{2014}).

\bibitem[{\citenamefont{Doi and Edwards}(1988)}]{doi1988theory}
\bibinfo{author}{\bibfnamefont{M.}~\bibnamefont{Doi}} \bibnamefont{and}
  \bibinfo{author}{\bibfnamefont{S.~F.} \bibnamefont{Edwards}},
  \emph{\bibinfo{title}{The Theory of Polymer Dynamics}}
  (\bibinfo{publisher}{Oxford University Press}, \bibinfo{year}{1988}).

\bibitem[{\citenamefont{Hansen and McDonald}(1990)}]{hansen1990theory}
\bibinfo{author}{\bibfnamefont{J.-P.} \bibnamefont{Hansen}} \bibnamefont{and}
  \bibinfo{author}{\bibfnamefont{I.~R.} \bibnamefont{McDonald}},
  \emph{\bibinfo{title}{Theory of Simple Liquids}}
  (\bibinfo{publisher}{Elsevier}, \bibinfo{year}{1990}).

\bibitem[{\citenamefont{Singh}(1991)}]{singh1991density}
\bibinfo{author}{\bibfnamefont{Y.}~\bibnamefont{Singh}},
  \bibinfo{journal}{Phys. Rep.} \textbf{\bibinfo{volume}{207}},
  \bibinfo{pages}{351} (\bibinfo{year}{1991}).

\bibitem[{\citenamefont{Evans}(1992)}]{evans1992density}
\bibinfo{author}{\bibfnamefont{R.}~\bibnamefont{Evans}}, in
  \emph{\bibinfo{booktitle}{Fundamentals of Inhomogeneous Fluids}}, edited by
  \bibinfo{editor}{\bibfnamefont{D.}~\bibnamefont{Henderson}}
  (\bibinfo{publisher}{Marcel Dekker}, \bibinfo{year}{1992}), pp.
  \bibinfo{pages}{85--176}.

\bibitem[{\citenamefont{Evans}(2010)}]{evans2010density}
\bibinfo{author}{\bibfnamefont{R.}~\bibnamefont{Evans}}, in
  \emph{\bibinfo{booktitle}{Lecture Notes 3rd Warsaw School of Statistical
  Physics}}, edited by
  \bibinfo{editor}{\bibfnamefont{B.}~\bibnamefont{Cichocki}},
  \bibinfo{editor}{\bibfnamefont{M.}~\bibnamefont{Napi\'{o}rkowski}},
  \bibnamefont{and} \bibinfo{editor}{\bibfnamefont{J.}~\bibnamefont{Piasecki}}
  (\bibinfo{publisher}{Warsaw University Press}, \bibinfo{year}{2010}), pp.
  \bibinfo{pages}{43--85}.

\bibitem[{\citenamefont{L{\"o}wen}(2010)}]{lowen2010density}
\bibinfo{author}{\bibfnamefont{H.}~\bibnamefont{L{\"o}wen}}, in
  \emph{\bibinfo{booktitle}{Lecture Notes 3rd Warsaw School of Statistical
  Physics}}, edited by
  \bibinfo{editor}{\bibfnamefont{B.}~\bibnamefont{Cichocki}},
  \bibinfo{editor}{\bibfnamefont{M.}~\bibnamefont{Napi\'{o}rkowski}},
  \bibnamefont{and} \bibinfo{editor}{\bibfnamefont{J.}~\bibnamefont{Piasecki}}
  (\bibinfo{publisher}{Warsaw University Press}, \bibinfo{year}{2010}), pp.
  \bibinfo{pages}{87--121}.

\bibitem[{\citenamefont{Risken}(1984)}]{risken1984fokker}
\bibinfo{author}{\bibfnamefont{H.}\hspace{0.06cm}\bibnamefont{Risken}},\hspace{0.05cm}\emph{\bibinfo{title}{The\hspace{0.06cm}Fokker-Planck\hspace{0.06cm}Equation}}\hspace{0.06cm}(\bibinfo{publisher}{Springer},\hspace{0.05cm}\bibinfo{year}{1984}).

\bibitem[{\citenamefont{Jiang et~al.}(2010)\citenamefont{Jiang, Yoshinaga, and
  Sano}}]{jiang2010active}
\bibinfo{author}{\bibfnamefont{H.-R.} \bibnamefont{Jiang}},
  \bibinfo{author}{\bibfnamefont{N.}~\bibnamefont{Yoshinaga}},
  \bibnamefont{and} \bibinfo{author}{\bibfnamefont{M.}~\bibnamefont{Sano}},
  \bibinfo{journal}{Phys. Rev. Lett.} \textbf{\bibinfo{volume}{105}},
  \bibinfo{pages}{268302} (\bibinfo{year}{2010}).

\bibitem[{\citenamefont{Volpe et~al.}(2011)\citenamefont{Volpe, Buttinoni,
  Vogt, K{\"u}mmerer, and Bechinger}}]{volpe2011microswimmers}
\bibinfo{author}{\bibfnamefont{G.}~\bibnamefont{Volpe}},
  \bibinfo{author}{\bibfnamefont{I.}~\bibnamefont{Buttinoni}},
  \bibinfo{author}{\bibfnamefont{D.}~\bibnamefont{Vogt}},
  \bibinfo{author}{\bibfnamefont{H.-J.} \bibnamefont{K{\"u}mmerer}},
  \bibnamefont{and}
  \bibinfo{author}{\bibfnamefont{C.}~\bibnamefont{Bechinger}},
  \bibinfo{journal}{Soft Matter} \textbf{\bibinfo{volume}{7}},
  \bibinfo{pages}{8810} (\bibinfo{year}{2011}).

\bibitem[{\citenamefont{Buttinoni et~al.}(2012)\citenamefont{Buttinoni, Volpe,
  K{\"u}mmel, Volpe, and Bechinger}}]{buttinoni2012active}
\bibinfo{author}{\bibfnamefont{I.}~\bibnamefont{Buttinoni}},
  \bibinfo{author}{\bibfnamefont{G.}~\bibnamefont{Volpe}},
  \bibinfo{author}{\bibfnamefont{F.}~\bibnamefont{K{\"u}mmel}},
  \bibinfo{author}{\bibfnamefont{G.}~\bibnamefont{Volpe}}, \bibnamefont{and}
  \bibinfo{author}{\bibfnamefont{C.}~\bibnamefont{Bechinger}},
  \bibinfo{journal}{J. Phys.: Condens. Matter} \textbf{\bibinfo{volume}{24}},
  \bibinfo{pages}{284129} (\bibinfo{year}{2012}).

\bibitem[{\citenamefont{Babel et~al.}(2014)\citenamefont{Babel, ten Hagen, and
  L{\"o}wen}}]{babel2014swimming}
\bibinfo{author}{\bibfnamefont{S.}~\bibnamefont{Babel}},
  \bibinfo{author}{\bibfnamefont{B.}~\bibnamefont{ten Hagen}},
  \bibnamefont{and}
  \bibinfo{author}{\bibfnamefont{H.}~\bibnamefont{L{\"o}wen}},
  \bibinfo{journal}{J. Stat. Mech.} \textbf{\bibinfo{volume}{2014}},
  \bibinfo{pages}{P02011} (\bibinfo{year}{2014}).

\bibitem[{\citenamefont{Menzel}(2015{\natexlab{b}})}]{menzel2015focusing}
\bibinfo{author}{\bibfnamefont{A.\hspace{0.06cm}M.}\hspace{0.06cm}\bibnamefont{Menzel}},\hspace{0.07cm}\bibinfo{journal}{EPL\hspace{0.07cm}(Europhys.\hspace{0.07cm}Lett.)}\hspace{0.07cm}\textbf{\bibinfo{volume}{110}},\hspace{0.07cm}\bibinfo{pages}{38005}\hspace{0.07cm}(\bibinfo{year}{2015}{\natexlab{b}}).

\bibitem[{\citenamefont{Tailleur and Cates}(2009)}]{tailleur2009sedimentation}
\bibinfo{author}{\bibfnamefont{J.}~\bibnamefont{Tailleur}} \bibnamefont{and}
  \bibinfo{author}{\bibfnamefont{M.~E.} \bibnamefont{Cates}},
  \bibinfo{journal}{EPL (Europhys. Lett.)} \textbf{\bibinfo{volume}{86}},
  \bibinfo{pages}{60002} (\bibinfo{year}{2009}).

\bibitem[{\citenamefont{Enculescu and Stark}(2011)}]{enculescu2011active}
\bibinfo{author}{\bibfnamefont{M.}~\bibnamefont{Enculescu}} \bibnamefont{and}
  \bibinfo{author}{\bibfnamefont{H.}~\bibnamefont{Stark}},
  \bibinfo{journal}{Phys. Rev. Lett.} \textbf{\bibinfo{volume}{107}},
  \bibinfo{pages}{058301} (\bibinfo{year}{2011}).

\bibitem[{\citenamefont{Pototsky and Stark}(2012)}]{pototsky2012active}
\bibinfo{author}{\bibfnamefont{A.}~\bibnamefont{Pototsky}} \bibnamefont{and}
  \bibinfo{author}{\bibfnamefont{H.}~\bibnamefont{Stark}},
  \bibinfo{journal}{EPL (Europhys. Lett.)} \textbf{\bibinfo{volume}{98}},
  \bibinfo{pages}{50004} (\bibinfo{year}{2012}).

\bibitem[{\citenamefont{Ono et~al.}(2002)\citenamefont{Ono, O'Hern, Durian,
  Langer, Liu, and Nagel}}]{ono2002effective}
\bibinfo{author}{\bibfnamefont{I.~K.} \bibnamefont{Ono}},
  \bibinfo{author}{\bibfnamefont{C.~S.} \bibnamefont{O'Hern}},
  \bibinfo{author}{\bibfnamefont{D.~J.} \bibnamefont{Durian}},
  \bibinfo{author}{\bibfnamefont{S.~A.} \bibnamefont{Langer}},
  \bibinfo{author}{\bibfnamefont{A.~J.} \bibnamefont{Liu}}, \bibnamefont{and}
  \bibinfo{author}{\bibfnamefont{S.~R.} \bibnamefont{Nagel}},
  \bibinfo{journal}{Phys. Rev. Lett.} \textbf{\bibinfo{volume}{89}},
  \bibinfo{pages}{095703} (\bibinfo{year}{2002}).

\bibitem[{\citenamefont{Seifert}(2012)}]{seifert2012stochastic}
\bibinfo{author}{\bibfnamefont{U.}~\bibnamefont{Seifert}},
  \bibinfo{journal}{Rep. Prog. Phys.} \textbf{\bibinfo{volume}{75}},
  \bibinfo{pages}{126001} (\bibinfo{year}{2012}).

\bibitem[{\citenamefont{Wensink et~al.}(2013)\citenamefont{Wensink, L{\"o}wen,
  Marechal, H{\"a}rtel, Wittkowski, Zimmermann, Kaiser, and
  Menzel}}]{wensink2013differently}
\bibinfo{author}{\bibfnamefont{H.~H.} \bibnamefont{Wensink}},
  \bibinfo{author}{\bibfnamefont{H.}~\bibnamefont{L{\"o}wen}},
  \bibinfo{author}{\bibfnamefont{M.}~\bibnamefont{Marechal}},
  \bibinfo{author}{\bibfnamefont{A.}~\bibnamefont{H{\"a}rtel}},
  \bibinfo{author}{\bibfnamefont{R.}~\bibnamefont{Wittkowski}},
  \bibinfo{author}{\bibfnamefont{U.}~\bibnamefont{Zimmermann}},
  \bibinfo{author}{\bibfnamefont{A.}~\bibnamefont{Kaiser}}, \bibnamefont{and}
  \bibinfo{author}{\bibfnamefont{A.~M.} \bibnamefont{Menzel}},
  \bibinfo{journal}{Eur. Phys. J. Special Topics}
  \textbf{\bibinfo{volume}{222}}, \bibinfo{pages}{3023} (\bibinfo{year}{2013}).

\bibitem[{\citenamefont{Baraban et~al.}(2012)\citenamefont{Baraban, Makarov,
  Streubel, M\"onch, Grimm, Sanchez, and Schmidt}}]{baraban2012catalytic}
\bibinfo{author}{\bibfnamefont{L.}~\bibnamefont{Baraban}},
  \bibinfo{author}{\bibfnamefont{D.}~\bibnamefont{Makarov}},
  \bibinfo{author}{\bibfnamefont{R.}~\bibnamefont{Streubel}},
  \bibinfo{author}{\bibfnamefont{I.}~\bibnamefont{M\"onch}},
  \bibinfo{author}{\bibfnamefont{D.}~\bibnamefont{Grimm}},
  \bibinfo{author}{\bibfnamefont{S.}~\bibnamefont{Sanchez}}, \bibnamefont{and}
  \bibinfo{author}{\bibfnamefont{O.~G.} \bibnamefont{Schmidt}},
  \bibinfo{journal}{ACS Nano} \textbf{\bibinfo{volume}{6}},
  \bibinfo{pages}{3383} (\bibinfo{year}{2012}).

\end{thebibliography}

\end{document}